\documentclass{aa}  

\usepackage{graphicx}
\usepackage{txfonts}
\usepackage{xcolor}
\newcommand{\cmnt}[1]{}

\newcommand{\porb}{$P_\mathrm{orb}$}

\newcommand{\inp}{{\tt input}}
\newcommand{\beer}{{\tt beer}}
\newcommand{\qmin}{{\tt qmin}}
\newcommand{\mmmr}{{\tt mmmr}}
\newcommand{\aell}{$A_{\rm ell}$}

\newcommand{\fbh}{$f_{\rm BH}$}
\newcommand{\sbar}{$\Bar{S}'$}

\newcommand{\mpia}{1}
\newcommand{\tlv}{2}
\newcommand{\barilan}{3}
\newcommand{\edin}{4}
\newcommand{\caltech}{5}
\newcommand{\ing}{6}
\newcommand{\sheff}{7}
\newcommand{\warw}{8}

\defcitealias{Green2023}{Paper I}

\begin{document}

   \title{An upper limit on the frequency of short-period black hole companions to Sun-like stars}


   \author{
Matthew J. Green\inst{\mpia},
Yoav Ziv\inst{\tlv},
Hans-Walter Rix\inst{\mpia},
Dan Maoz\inst{\tlv},
Ikram Hamoudy\inst{\barilan},
Tsevi Mazeh\inst{\tlv},
Simchon Faigler\inst{\tlv},
Marco C. Lam\inst{\edin},
Kareem El-Badry\inst{\caltech},
George Hume\inst{\ing,\sheff},
James Munday\inst{\warw},
and Paige Yarker\inst{\ing,\sheff}. 
          }

\institute{Max-Planck-Institut f\"{u}r Astronomie, K\"{o}nigstuhl 17, D-69117 Heidelberg, Germany\\ \email{mjgreenastro@gmail.com}
\and
School of Physics and Astronomy, Tel-Aviv University, Tel-Aviv 6997801, Israel
\and
Department of Physics, Bar-Ilan University, Ramat Gan 5290002, Israel
\and
Institute for Astronomy, University of Edinburgh, Royal Observatory of Edinburgh, Blackford Hill, Edinburgh EH9 3HJ, UK
\and
California Institute of Technology, 1200 E California Blvd, Pasadena, CA 91125, United States
\and
Isaac Newton Group of Telescopes, Apartado de Correos 368, E-38700 Santa Cruz de La Palma, Spain
\and
Department of Physics and Astronomy, University of Sheffield, Sheffield, S3 7RH, UK
\and
Astronomy and Astrophysics Group, Department of Physics, University of Warwick, Coventry, CV4 7AL, UK
             }

   \date{Received December 2, 2024}

 
\abstract
{Stellar-mass black holes descend from high-mass stars, most of which had stellar binary companions. 
However, the number of those binary systems that survive the binary evolution and black hole formation is uncertain by multiple orders of magnitude.
The survival rate is particularly uncertain for massive stars with low-mass companions, which are thought to be the progenitors of most black hole X-ray binaries.
We present a search for close black hole companions (orbital period $\lesssim 3$\,days, equivalent to separation $\lesssim 20 R_\odot$) to AFGK-type stars in \textit{TESS}; that is, the non-accreting counterparts to and progenitors of low-mass X-ray binaries. Such black holes can be detected by the tidally induced ellipsoidal deformation of the visible star, and the ensuing photometric light curve variations. From an initial sample of $4.7\times10^6$ \textit{TESS} stars, we have selected 457 candidate ellipsoidal variables with large mass ratios. 
However, after spectroscopic follow-up of 250 of them, none so far are consistent with a close black hole companion.
On the basis of this non-detection, we determine (with $2\sigma$ confidence) that fewer than one in $10^5$ solar-type stars in the solar neighbourhood hosts a short-period black hole companion.
This upper limit is in tension with a number of `optimistic' population models in the literature that predict short-period black hole companions around one in $\sim 10^{4-5}$ stars. 
Our limit is still consistent with other models that predict only a few in $\sim 10^{7-8}$.   
}


\authorrunning{M. J. Green et al.}

   \maketitle
%

\section{Introduction}

Isolated stars with masses $\gtrsim 15$--$25 M_\odot$ are believed to end their lives as black holes, which implies that $\sim 10^8$ stellar-mass black holes exist in our Galaxy \citep[e.g.][]{Olejak2020}.
However, the majority of these massive precursor stars were not formed in isolation, but rather had at least one stellar companion, and approximately half have a companion on a sufficiently close orbit that the two stars will interact within their lifetimes \citep{Sana2012,Moe2017}. 
Such interactions, including common envelope evolution or stable mass transfer, will affect the evolution of the individual stars and orbital properties in ways that are complex and not well understood \citep[e.g.][]{Mandel2022}.
In addition, mass loss during a massive star's evolution or in a supernova explosion will affect the orbital parameters and may even disrupt the binary completely.
Due to these uncertainties and their sensitivity to assumed parameters, predictions for the number of surviving binaries with one black hole component in our Galaxy vary by more than four orders of magnitude \citep[e.g.][]{Mashian2017,Breivik2017,Shao2019,Shikauchi2023}.

Free-floating black holes are difficult to find and study, with only one high-confidence example found so far \citep[discovered via gravitational microlensing;][]{Lam2022,Sahu2022,Mroz2022,Lam2023}. 
The majority of detected stellar-mass black holes have been found in binary systems, either as extragalactic gravitational wave sources consisting of two black holes \citep[e.g.][]{Abbott2023}\footnote{While the detection of gravitational waves has allowed for the component masses to be measured in a large number of double-black hole binary systems, the masses of these extragalactic sources appear to be systematically larger than the majority of known Galactic black holes, perhaps suggesting a different formation mechanism.}, or as mass-transferring binaries detected by their X-ray emission.
In the latter category, there are $\approx20$ X-ray binaries with dynamically confirmed black hole accretors and another $\approx50$ containing candidate black holes \citep[e.g.][]{Corral-Santana2016}.
These constitute the vast majority of known black holes in the Galaxy, but presumably represent only a fraction of the possible parameter space of black hole binary systems.

Significant research effort has been put into the search for non-accreting (`dormant')  black holes with luminous companions (BH-LCs), but to date only eight solid discoveries have been made.
Contemporary searches for BH-LCs tend to make use of one of three approaches, each of which is most sensitive to different ranges of orbital separations: astrometric orbital fitting \citep[e.g.][]{Shahaf2022,El-Badry2023b,El-Badry2023a}; spectroscopic searches for large radial velocity (RV) shifts in the luminous star \citep[e.g.][]{Giesers2018,Giesers2019,Mahy2022,Shenar2022}; and searches for photometric signatures of the tidal distortion of the luminous star \citep[ellipsoidal modulation; e.g.][]{Gomel2021c,Gomel2023,Rowan2021,Rowan2024,Kapusta2023}.
From all of these efforts, only a small number of dormant BH-LCs have been found: three detections of black holes with low-mass companions based on \textit{Gaia} astrometry \citep{El-Badry2023b,El-Badry2023a,Chakrabarti2023,Tanikawa2023,GaiaCollaboration2024}; two black holes with high-mass companions detected by spectroscopic surveys \citep{Mahy2022,Shenar2022}, of which one is in the Milky Way and one in the Large Magellanic Cloud; one mass-gap black hole candidate with a low-mass giant companion detected through \textit{Gaia} spectroscopic data \citep{Wang2024}; and two black holes with low-mass companions in the globular cluster NGC 3201, discovered with spectroscopic surveys \citep{Giesers2018,Giesers2019}\footnote{A number of candidate neutron star companions to luminous stars have also been found using similar methods \citep[e.g.][]{Mazeh2022,Geier2023,El-Badry2024a,El-Badry2024b}.}.
The orbital periods of field, non-accreting BH-LCs discovered so far range from approximately ten to several thousand days\footnote{One of the NGC 3201 systems has a period of 2.24 days, but we note that interactions in dense stellar environments tend to shrink the orbits of binary systems.}.
The current numbers of such discoveries seem to fall short of the hundreds predicted by most population models, but it is currently unclear whether the discrepancy is due to a true shortage of BH-LCs, or because of the complex and difficult-to-reproduce selection processes so far applied by \textit{Gaia}, combined with the fact that most spectroscopic surveys do not obtain enough RV epochs to constrain orbital periods and companion masses \citep[see the discussion by][]{El-Badry2024}. 
Based on a single detection, \citet{El-Badry2023a} estimated that a fraction of $\approx 4 \times 10^{-7}$ solar-type stars may have companions in the period range of 300--1000 days ($\approx 10^{-6}$ per host star per dex of orbital period, assuming a log-uniform period distribution), with significantly weaker constraints at shorter orbital periods to which astrometry is insensitive.

Given the paucity of true BH-LC discoveries, it is useful to set observational constraints on this population
by deriving upper limits on how common such objects are, based on a simple, well-defined, and large parent sample of 
stars with well-understood selection effects and detection efficiencies. The dependence of the frequency of BH-LC systems on orbital separation or various properties of the luminous stars can also be probed.
In this study, we  utilise one of the survey methods listed above, the search for ellipsoidal variability, focussing on the shortest periods among putative dormant black hole companions.
Ellipsoidal variability is the photometric signature of the tidal distortion of the photometrically dominant star in a binary system by the gravity of a close companion \citep[e.g.][]{Kopal1959,Morris1985,Morris1993,Faigler2011}.
If the luminous star is a main-sequence, solar-type star, and the selection process is sensitive to amplitudes $\approx 1$ part per thousand, then this method is sensitive to dark companions with orbital periods of \porb\ $\lesssim 3$\,days.
This is an interesting period range: it is only slightly longer than the typical periods of low-mass X-ray binaries, yet no non-accreting BH-LC has yet been found in this period range in the field \citep[although at least one candidate dormant neutron star has been claimed;][]{Mazeh2022}.

Searching for BH-LCs via the ellipsoidal light curve signature has seen significant interest in the last several years. 
\citet{Gomel2021a,Gomel2021b,Gomel2021c,Gomel2023} discussed the methodology extensively, and produced a list of candidate BH-LCs from \textit{Gaia} photometric data.
\citet{Rowan2021} and \citet{Kapusta2023} have applied similar methodologies to photometry from the All-Sky Automated Survey for SuperNovae (ASAS-SN) and the Optical Gravitational Lensing Experiment (OGLE).
A number of these candidates have been followed up \citep{Nagarajan2023,Kapusta2023,Rowan2024} and to date produced no likely BH-LC candidate -- making this the only one of the three search methods listed above to have not yet led to a BH-LC discovery. 
It may be that black hole companions at these short periods are rarer than at longer periods, but statistical assessments in both period ranges are needed before such a claim can be made.

The goal of this study is to place an upper limit on the space density of short-period (\porb\ $\lesssim 3$\,days) BH-LCs.
In order to constrain the space density of the underlying population after selecting candidates through a given selection method, it is necessary to understand the efficiency of the selection method as a function of the input physical parameters.
In a previous work \citep[][henceforth \citetalias{Green2023}]{Green2023}, we selected a sample of 15\,000 candidate ellipsoidal binary systems, primarily consisting of two main-sequence stars (MS stars, MS-MS binaries) in either a detached or contact configuration, based on their Transiting Exoplanet Survey Satellite (\textit{TESS}) photometry.
An advantage of that sample is that significant effort was put into understanding the efficiency of our selection algorithm.
In this work, we perform a search for BH-LCs hidden among that sample, show a number of non-detections using follow-up spectroscopy, and use the overall
non-detection to estimate an upper limit on the space density of BH-LCs that are accessible to this method.

In Section~\ref{sec:candidates}, we describe the candidate selection process. 
Sections~\ref{sec:observations} and \ref{sec:amplitudes} describe the follow-up observations undertaken and the data analysis performed to measure the RV amplitude, while Section~\ref{sec:gaia} describes the data retrieved from the \textit{Gaia} catalogue.
Section~\ref{sec:upper-lim} describes the estimation of an upper limit from these non-detections, and in Section~\ref{sec:discussion} we discuss the implications of this upper limit.









\section{Candidate selection}
\label{sec:candidates}

\subsection{Initial ellipsoidal selection with BEER}

We begin with the sample of ellipsoidal binary systems in \textit{TESS}, selected in \citetalias{Green2023}.
A full description of the selection process is given in that paper, but we provide here a brief overview.

In \citetalias{Green2023}, we processed full-frame image light curves of all targets from the first two years of the \textit{TESS} mission with magnitudes of $T < 13.5$ (approximately $G < 13$ for solar-type stars).
Those light curves have a cadence of 30\,min, and a typical length of 27\,days (although approximately a third of targets fell into multiple observing sectors and so have light curves over a longer timespan).
Before processing, a cut in absolute magnitude versus colour was applied to remove targets that were well above the main sequence, resulting in 4.7 million remaining targets.

These target light curves were then processed using the BEaming, Ellipsoidal modulation, and Reflection  ({\sc beer}) algorithm \citep{Mazeh2010,Mazeh2012,Faigler2011,Faigler2015a,Gomel2021a,Gomel2021b}, which fits for signatures of Doppler beaming, ellipsoidal modulation, and reflection.
Initially, each light curve was fit with a simple model of three sinusoids (assumed to be at the orbital period and its lowest two harmonics).
We then selected the light curves for which the ellipsoidal signature (frequency $2 / P_{\rm orb}$) and at least one other harmonic were significantly detected.
Subsequently, a physical model was converged on the measured amplitudes, and any target for which no physical solution was found was discarded. (We note that, even if a physical solution was found, it is not necessarily a unique solution.)
Finally, several cuts were applied in amplitude and period space in order to remove regions that are known to be dominated by non-binary contaminants (pulsators or rotating spotted stars).

We note that, although the {\sc beer} algorithm also fits for reflection and Doppler beaming, only the measured ellipsoidal amplitude is used in the following sections.
As is discussed in \citetalias{Green2023}, the measured Doppler beaming amplitude can be strongly affected by the presence of star spots, and we therefore consider it to be less reliable than the ellipsoidal modulation.
See Section~\ref{sec:positives} for a more detailed discussion of star spots.

This process reduced 4.7 million main-sequence targets to 15\,000 candidate ellipsoidal binary systems, with an estimated purity (rate of true positive binaries) of 80--90 percent and an estimated completeness of $28 \pm 3$ percent of all main-sequence-primary binary systems with \porb\ $\lesssim 3$\,days \citepalias{Green2023}.
Throughout this work, we refer to the initial 4.7 million targets as the \inp\ sample, and the 15\,000 ellipsoidal candidates as the \beer\ sample.

From this \beer\ sample of binary candidates, it was necessary to choose the subset of targets which are the most promising BH-LC candidates, rather than MS-MS binaries or non-binary contaminants.
We applied two parallel selection methods.
Both methods involved estimating a lower limit on the mass ratio $q = M_2 / M_1$ (where $M_1$ is the mass of the photometric primary star, and $M_2$ the photometric secondary), on the premise that, for a  main-sequence primary star,  $q > 1$ is only possible if the secondary star is a high-mass compact object.
The methods differed in how to calculate this lower limit. The first was a method based on the standard ellipsoidal mass function, which we refer to as the \qmin\ method.
The second method was based on the modified minimum mass ratio (MMMR) statistic proposed by \citet{Gomel2021b}, which we refer to as the \mmmr\ method.
Neither selection is perfect;  we 
demonstrate below that the \qmin\ method typically over-estimates $q$ at short orbital periods, while the \mmmr\ method is insensitive even to high-mass companions except at very short periods.
The BH-LC candidates selected by the two methods, which we describe in detail below, will be referred to as the \qmin\ and \mmmr\ samples, respectively.
A summary of the number of candidates remaining after each step in the selection processes can be found in Table~\ref{tab:candidates}.

\begin{table}
\caption{Number of candidates remaining after each step in the \qmin\ and \mmmr\ selection process.}
\begin{tabular}{lr}
\hline
Stage & Number of candidates\\
\hline
\inp\ & $4.7 \times 10^6$\\
\beer\ ellipsoidals & 15779\\
\qmin\ initial selection & 847\\
\quad ... not eclipsing & 411\\
\mmmr\ initial selection & 1501\\
\quad ... not eclipsing & 46\\
Total candidates & 457\\
\\
Observed, \qmin\ & 231\\
Observed, \mmmr\ & 19\\
Observed, total & 250\\
\hline
\end{tabular}
\label{tab:candidates}
\end{table}

\subsection{\qmin\ method}
\label{sec:qmin}

\begin{figure*}
\resizebox{\hsize}{!}
{\includegraphics{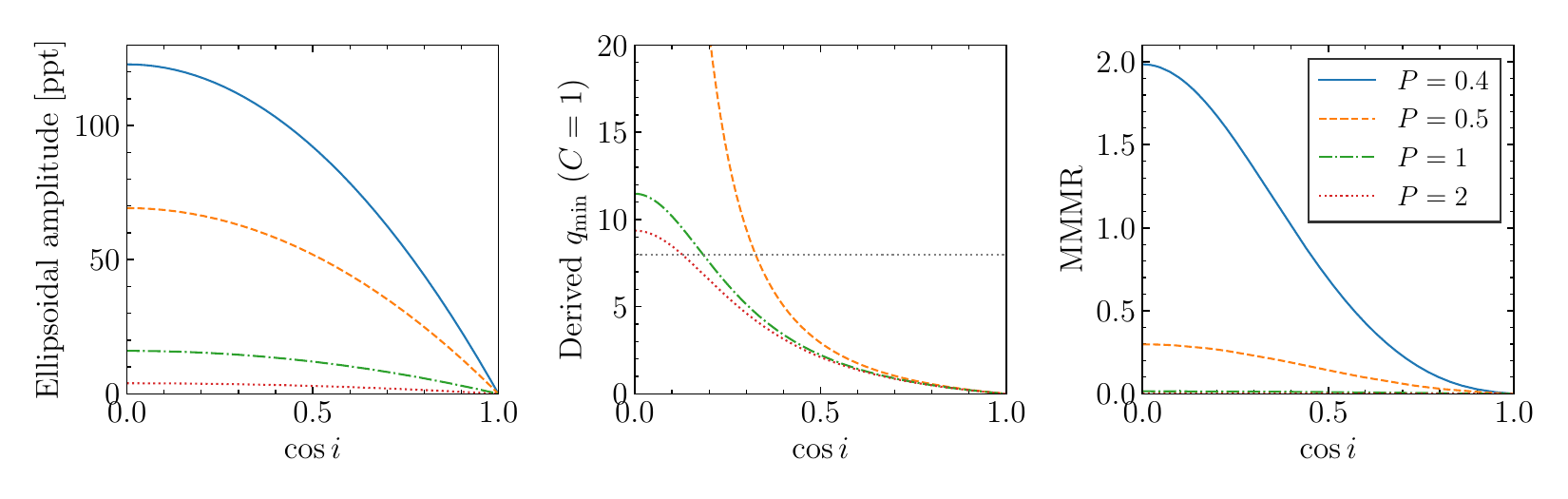}}
\caption{ \textit{Left:} Expected \aell\ for a $1\,M_\odot$ MS star with an $8 M_\odot$ dark companion (i.e. mass ratio $q=8$) at a range of orbital periods (denoted in days in the legend) as a function of orbital inclination $\cos i$ (0 is edge-on, 1 is face-on). Because the probability distribution of $\cos i$ is uniform for a randomly oriented orbit, every $y$ value plotted here is equally likely. We note that this binary system will overflow its Roche lobe at a period of $\approx 0.37\,$days.
\textit{Middle:} Derived values of $q_{\rm min}$ for the same binary systems, assuming perfect knowledge of $M_1$ and $R_1$. 
The differences between the curves in this panel come solely from neglecting of the \citet{Gomel2021b} correction factor, $C$, which depends on Roche lobe filling factor.
Neglecting $C$ also explains why values of $q_{\rm min} > 8$ are possible, as is discussed in the text. 
The dotted grey line shows the true value of $q$. We do not plot the \porb\,$= 0.4$\,day model here, as the large value of $C$ produces an unphysical value of $\mathcal{M}_{\rm ell} > 1$ and causes a breakdown of the $q_{\rm min}$ calculation. 
\textit{Right:} Derived values of MMMR for the same binary systems. We note that most binary configurations result in MMMR\,$< 1$, and even the \porb\,$= 0.4$\,day track produces MMMR\,$> 1$ for less than half of the range of $\cos i$.
}
\label{fig:method-tracks}
\end{figure*}

\begin{figure}
\centering
\includegraphics[width=\hsize]{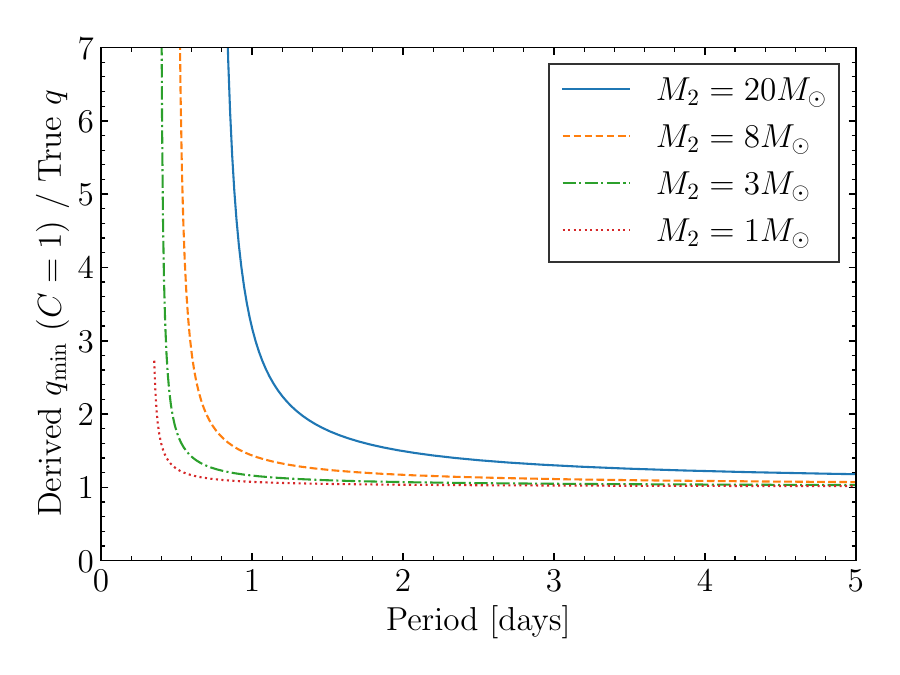}
\caption{Values of the maximum fraction (derived $q_{\rm min}$ / true $q$) by which $q_{\rm min}$ may be overestimated due to the unknown value of $C$. We assume here a Sun-like primary star and an edge-on inclination.
For \porb\,$\gtrsim 1$\,day, the overestimation is relatively minor unless the donor is unusually massive.
}
\label{fig:method-tracks-qmin}
\end{figure}

\begin{figure}
\centering
\includegraphics[width=\hsize]{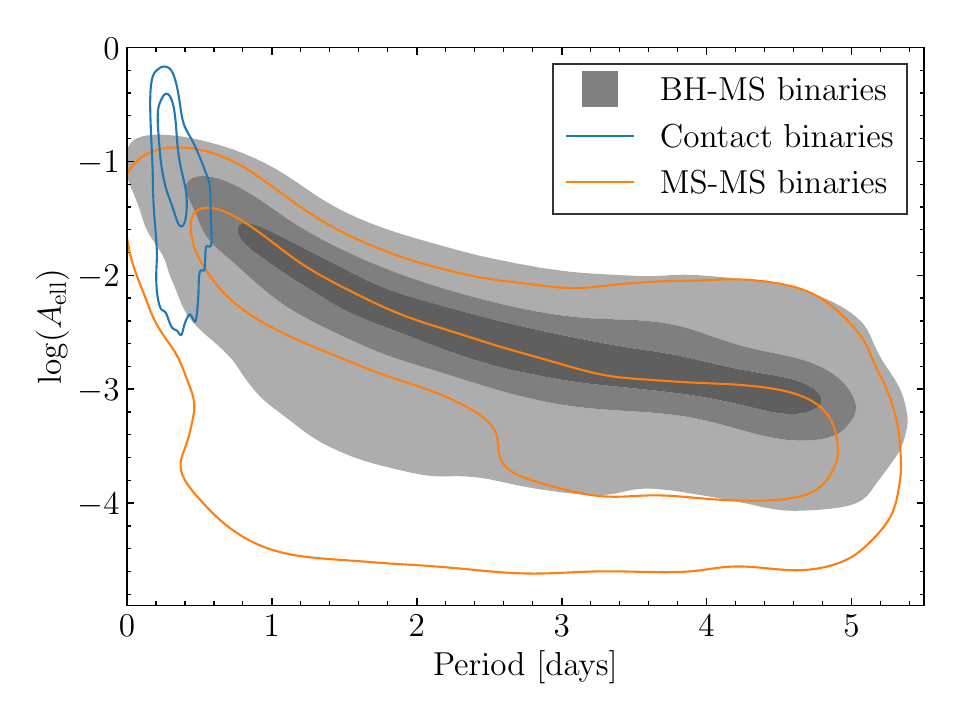}
\caption{Simulated \aell\ distributions for BH-LCs (black), detached MS-MS binaries (orange), and contact binaries (cyan). 
Contours outline the regions of constant density.
The simulation of BH-LCs is described in Section~\ref{sec:simulations}, while the other simulated populations are described in \citetalias{Green2023}.
BH-LCs typically have a somewhat larger mean amplitude than detached MS-MS, by between 0.25 to 0.5 dex depending on the period, but the amplitudes of contact binaries can be even larger.
}
\label{fig:simulation-amplitude}
\end{figure}

Ellipsoidal distortions in the primary stars result in light curve signals at several harmonics of the orbital period, but (in MS-MS or BH-LC binaries) the strongest signature is always\footnote{Always in the simple case where all variability is ellipsoidal. Other sources of variability such as star spots may introduce stronger variability at the orbital frequency, as is discussed in Section~\ref{sec:positives}.} at a frequency $2 / P_{\rm orb}$.
In general, for a detached ellipsoidal binary system, the amplitude of the signal is approximated \citep{Morris1993,Faigler2011,Gomel2021b} by
\begin{equation}
\begin{split}
A_{\rm ell} \approx  &\left( f_1 \alpha_\text{1} C_{1} \frac{M_2}{M_1} \left(\frac{R_1}{a}\right)^3 + f_2 \alpha_\text{2} C_{2} \frac{M_1}{M_2} \left(\frac{R_2}{a}\right)^3 \right) \sin^2 i \\
\approx  &13400 \sin^2 i \left(\frac{M_1 + M_2}{M_\odot}\right)^{-1} \left(\frac{P_\text{orb}}{\text{day}}\right)^{-2} \\
&\times \left(f_1 \alpha_\text{1} C_{1} \frac{M_2}{M_1} \left(\frac{R_1}{R_\odot}\right)^3 + f_2 \alpha_\text{2} C_{2} \frac{M_1}{M_2} \left(\frac{R_2}{R_\odot}\right)^3  \right) \text{ppm},
\label{eq:ell2}
\end{split}
\end{equation}
where ppm is parts per million, $M_1$ and $M_2$ refer to the masses of the component stars, $R_1$ and $R_2$ are their radii, $f_1$ and $f_2$ are their relative fluxes (such that $f_1 + f_2 = 1$), $a$ is the orbital semi-major axis, \porb\ is the orbital period, $i$ is the orbital inclination, $\alpha_1 \approx \alpha_2 \approx 1.3$ are constants that depend weakly on limb darkening and gravity darkening, and $C_1$ and $C_2$ are correction factors introduced by \citet{Gomel2021b} that depend on $q$ and the Roche-lobe filling factor. 
The factors $C_1$ and $C_2$ typically have values between 1 and 1.5, and are necessary when the stars are close to filling their Roche lobes.
We note that we adopt here a convention in which \aell\ is positive, which is a change from \citetalias{Green2023}.

Henceforth, we make the assumption that all of the light in the binary comes from the primary star ($f_1 = 1, f_2 = 0$).
If both stars lie on the main sequence, Equation~\ref{eq:ell2} will be a good approximation of the true \aell\ if $q = M_2 / M_1 \lesssim 0.3$ or $q \gtrsim 0.8$, while it overestimates \aell~ at the level of $\approx$ 5--10 percent if $q \approx 0.5$ and leads to an eventual overestimation of $q_{\rm min}$ by a similar fraction -- not large enough to bring any detached systems from $q \approx 0.5$ into our $q_{\rm min} > 1$ sample.

Further to the above assumption, we are also forced to neglect the correction factor, $C_1$, as calculating the correction factor would require knowledge of $q$, and so
we assume $C_1 = 1$. 
For binaries in which the primary star is close to filling its Roche lobe, this can lead to a substantial overestimate of $q_{\rm min}$, as is shown in Fig.~\ref{fig:method-tracks}, 
which illustrates the estimated lower limits on $q$ for several example binaries.
At \porb\,$\gtrsim 1$\,day, because the Roche lobes are larger, the overestimation of $q_{\rm min}$ is only substantial if $q \gtrsim 1$ (Fig.~\ref{fig:method-tracks-qmin}).

Under these assumptions, we can define the ellipsoidal mass function (by analogy with the spectroscopic mass function) as
\begin{equation}
\mathcal{M}_{\rm ell}  = \sin^2 i \frac{q}{q+1}
= \frac{1}{13400 \,\alpha_1} \frac{M_1}{M_\odot} \left(\frac{R_1}{R_\odot}\right)^{-3} \left(\frac{P_\text{orb}}{\text{day}}\right)^{2}  A_{\rm ell},
\end{equation}
where \aell\ is expressed in units of parts per million as in Equation~\ref{eq:ell2}.
We can derive a lower limit on $q$ by setting $\sin i = 1$: 
\begin{equation}
q_{\rm min} =  \frac{\mathcal{M}_{\rm ell}}{1-\mathcal{M}_{\rm ell}} .
\end{equation}
If we assume $\alpha_1 = 1.3$, and we have good estimates of $M_1$ and $R_1$, then $q_{\rm min}$ can be estimated from just the ellipsoidal amplitude and period.
Approximately 10\% of detached binaries and 20\% of contact binaries with \porb\,$< 1$\,day become contaminants using the \qmin\ method due to the neglection of $C_1$, while $\approx 5$\% of detached systems become contaminants with $1 <$\porb$< 2$\,days. 
Neglecting the correction factor, $C_1$, can therefore introduce a significant source of false positives, especially at \porb\,$\lesssim 1$\,day.

Any ellipsoidal binary system for which $q_{\rm min} > 1$ should, in principle, have a high-mass, dark companion, as long as the primary star is truly on the main sequence.
Equation~\ref{eq:ell2} relies on the assumption that the stars are detached, so contact binaries (which tend to have larger amplitudes but otherwise similar light curves) are an important source of false positives, as others have already found \citep[e.g.][]{Nagarajan2023,Kapusta2023}.
This can also be seen in Fig.~\ref{fig:simulation-amplitude}, showing the amplitudes of different types of ellipsoidal binary systems.
As the vast majority of FGK-type contact binaries have \porb~$< 1$\,day, we apply an additional cut to the \qmin\ sample of \porb~$> 1$\,day to avoid the contaminated region of parameter space.
This also reduces the effect of our $C_1 = 1$ assumption, discussed above.
It should be noted that star spots can also introduce false positives by increasing the measured \aell\ (discussed further in Section~\ref{sec:positives}).

\begin{figure*}
\centering
\includegraphics[width=\hsize]{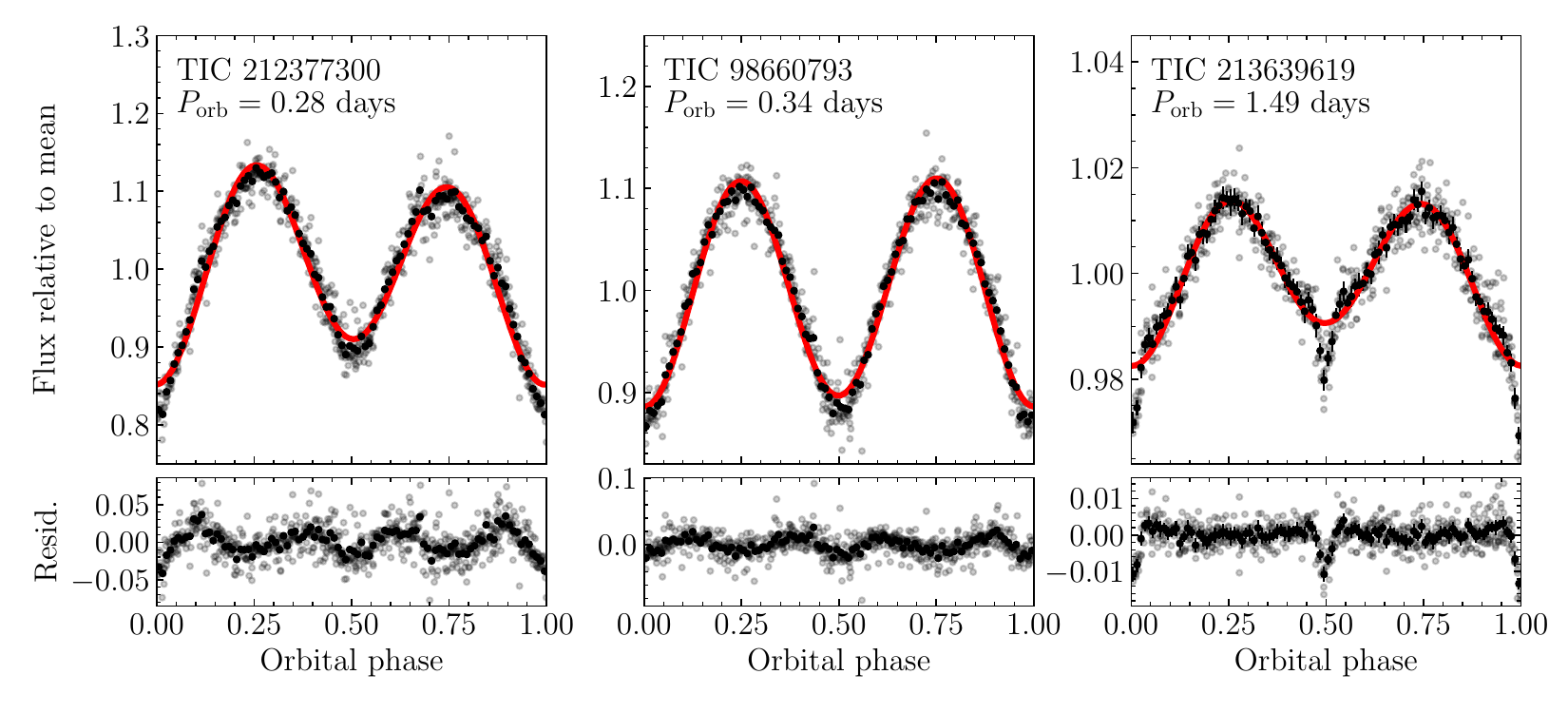}
\caption{Several example light curves that were identified as showing eclipses. 
Grey points show the individual \textit{TESS} data, black points are phase-binned data, and the red line shows the best-fit ellipsoidal model.
Eclipse features may be shallow, but can be identified as a departure from the ellipsoidal model at orbital phases 0 and 0.5.
}
\label{fig:eclipsers}
\end{figure*}

At short orbital periods, a significant fraction of binary systems eclipse.
This can be used to remove a fraction of false positives from our sample of BH-LC candidates, as black holes do not cause eclipses and are not dimmed by them.
We performed a by-eye inspection of the light curves of every target in the \qmin\ sample, checking both the light curve itself and the residuals (after subtracting the best-fit ellipsoidal model) for eclipse features at phases 0 or 0.5.
We first trained our eyes using 1000 synthetic light curves of detached binaries generated using the {\tt ellc} package, including both eclipsing and non-eclipsing targets, and found that we were able to reliably identify eclipse depths of $\gtrsim 1$ percent.
We then looked through all light curves of our targets in the same way and removed any that showed eclipses.
Several example light curves of real eclipsing targets that were removed in this way are shown in Fig.~\ref{fig:eclipsers}.
This removed 436 targets.

\begin{figure}
\centering
\includegraphics[width=\hsize]{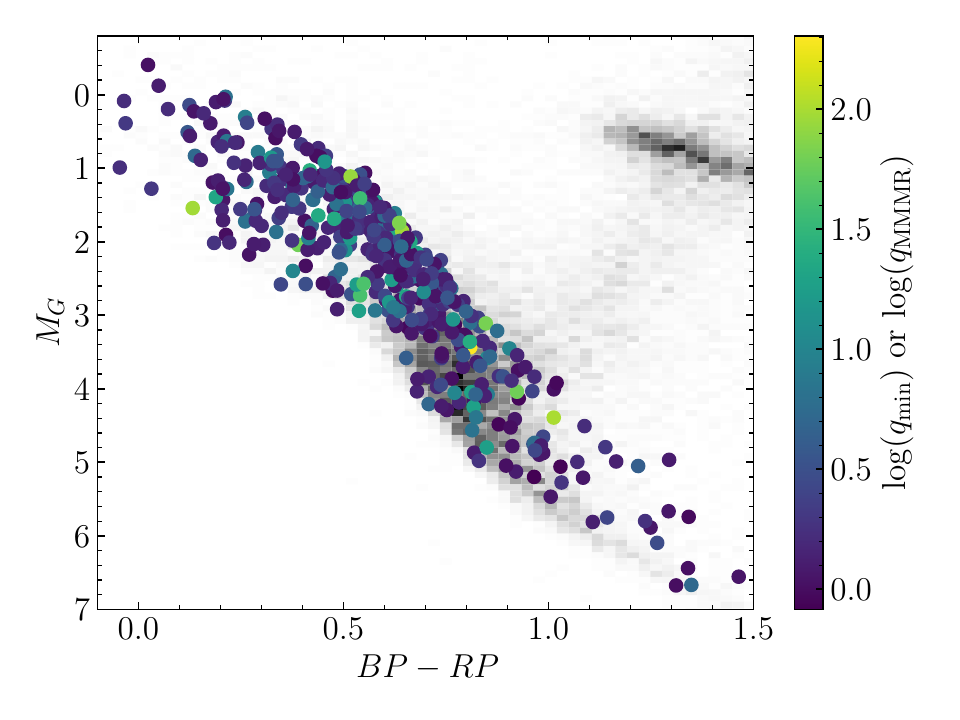}
\caption{Colour-magnitude diagram of our targets (coloured circles) compared to a magnitude-limited sample of stars (grey two-dimensional histogram). Our targets are coloured according to their ellipsoidal-implied $q_{\rm min}$ for targets with $P_{\rm orb} > 1$\,day (Section~\ref{sec:qmin}), or by MMMR for targets with $P_{\rm orb} < 1$\,day (Section~\ref{sec:mmmr}).
Many targets are somewhat elevated above the main sequence, but it is unclear whether this is due to light from the secondary star or due to inflation of the primary radius (we note that the latter is selected for by the ellipsoidal selection method).
}
\label{fig:cmd}
\end{figure}

The limitations of the \qmin\ approach have been discussed by \citet{Gomel2021b}. An important one, in our context, is that it relies heavily on accurate primary stellar masses and radii, which are not available for all of our targets.
We note that masses and radii estimated based on colour-magnitude position, assuming single stellar models, are not ideal for our purposes (as the most common kinds of false positives will contain light from two stars which may particularly bias radius estimates).
As was previously done in \citetalias{Green2023}, we estimate the primary stellar masses and radii from the effective temperatures tabulated in the \textit{TESS} Input Catalogue \citep[TIC, version 8;][]{Stassun2019}. 
We favour these temperature estimates as they depend only on colour, which is less strongly affected by the light of a main-sequence companion than absolute magnitude.
We then assume that the primary stars are on the main sequence and interpolate from the temperature to find mass and radius using the \citet{Pecaut2013} tables.
This assumption is not ideal, as the ellipsoidal selection method has an innate selection in favour of binaries with inflated primary stars.
Because of this limitation, the most common class of contaminants in the \qmin\ sample (at periods $> 1$\,day) are detached MS-MS binaries in which the primary is slightly inflated.
In Fig.~\ref{fig:cmd}, we plot a colour-magnitude diagram of our selected targets, showing that many are significantly elevated above the main sequence (especially at higher masses or $BP-RP \lesssim 0.6$).
Before obtaining further data, it is unclear for any given target whether this elevation is the result of light from a companion, inflation of the primary star, or a combination of both.
At the end of this selection process, 411 targets remained in the \qmin\ sample of candidate BH-LCs.

\subsection{Modified minimum mass ratio method}
\label{sec:mmmr}

\begin{figure}
\centering
\includegraphics[width=\hsize]{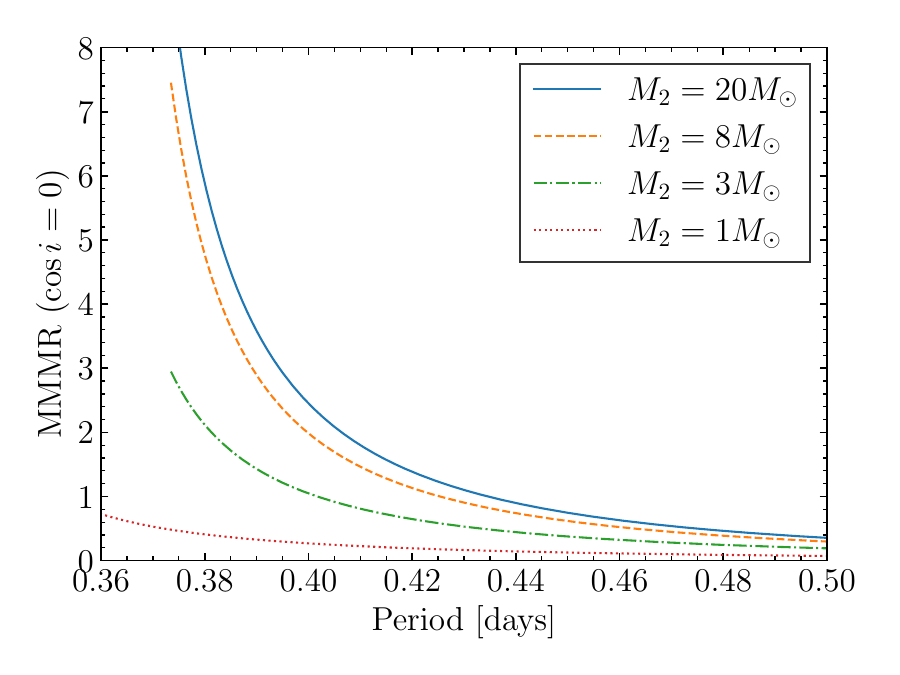}
\caption{Values of MMMR for a range of secondary masses, assuming a 1\,$M_\odot$ primary star and an edge-on inclination. Tracks are plotted for the period range in which neither star fills its Roche lobe. Even for unusually high-mass secondaries, MMMR\,$> 1$ is only possible for a Sun-like primary star over a narrow range of periods $0.37 \lesssim$\,\porb\,$\lesssim 0.44$\,days.
}
\label{fig:method-tracks-mmmr}
\end{figure}

We based our second approach on the MMMR suggested by \citet{Gomel2021b}, which was also used in the selection of the \textit{Gaia} ellipsoidal sample \citep{Gomel2023}.
In the MMMR approach, one assumes both that $\sin i = 1$ as in the previous section, and also that the primary star fills its Roche lobe, giving a stricter lower limit on $q$.
This approach bypasses the need for reliable primary stellar masses and radii.
The MMMR is a monotonic function of \aell, and can be found by first expressing \aell\ as a function of the Roche lobe filling factor, $F = R_1 / R_L$, where $R_L$ is the radius of the primary Roche lobe.
This gives from Equation~\ref{eq:ell2}:
\begin{equation}
A_{\rm ell} = \alpha_1 \, F^3 \, E(q)^3 \, q \, \sin^2 i 
 \, C_1(F,q),
\end{equation}
where $E(q) = R_L / a$ is the ratio of the primary Roche lobe radius to the orbital semimajor axis, which can be calculated using the approximation of \citet{Eggleton1983}.
As in the previous section, we have assumed that the primary star dominates all flux from the target ($f_1 = 1, f_2 = 0$).
After setting $\sin i = F = 1$, we can solve this equation numerically for $q$ to find the MMMR.

As with the \qmin\ selection method, the \mmmr\ selection method relies on the fact that the unseen, higher-mass companion to a detached main-sequence star must be a compact object.
The advantage of the \mmmr\ method is that it does not depend on reliable stellar mass and radius measurements. 
The disadvantage of the \mmmr\ method is that it is much stricter than \qmin, and can even remove many or most true BH-LCs from a sample.
As an example, for a $<20 M_\odot$ black hole around a Sun-like star, there is a very narrow period range ($0.37 \lesssim$\,\porb\,$\lesssim 0.44$\,days) in which MMMR\,$> 1$ is possible and the system remains detached, and even at such short periods the binary may only have MMMR\,$> 1$ for a minority of orbital inclinations (see Figs.~\ref{fig:method-tracks} and \ref{fig:method-tracks-mmmr}).
In other words, the \mmmr\ selection method will only detect a minority of all BH-LCs at orbital periods $\lesssim 0.5$\,days, and essentially none at longer orbital periods.

Therefore, it is not feasible to apply to the  \mmmr\ method the same period cut that we applied in the case of the \qmin\ method, and hence contact binaries remain the most significant source of contamination in \mmmr\ selection.
Several methods have been suggested in the literature to distinguish contact binaries from detached systems at the same orbital periods, including cuts in colour-period space \citep[e.g.][]{Rucinski2002}; the `morphology parameter', that quantitatively describes the smoothness of the light curve \citep[e.g.][]{Prsa2011,Prsa2022,Kirk2016}; and our own method, that involved cuts in amplitude-period space \citepalias{Green2023}.
We investigated several of these avenues but did not find any of them to work well. For example, there was significant disagreement between the systems highlighted as contact binaries by different methods. 
We note that \citet{Pesta2023,Pesta2024} have found success isolating contact binaries from the detached population in an automated way, using either a combination of morphology and colour information or light curve-based principal component analysis. Unfortunately their papers were released too late for us to implement their method in our target selection, but this approach may represent a useful route for future works in this area.

Instead, we found that the same by-eye search for eclipses in the light curve that we implemented in the \qmin\ section was effective at removing contact binaries. 
We generated a set of 40\,000 simulated light curves of contact binaries using the {\sc phoebe} package \citep{Prsa2005}.
After passing these through the {\tt beer} and {\tt mmmr} selection processes described above, around 1000 remained, which we examined by eye.
As in Section~\ref{sec:qmin}, we found that we were able to identify eclipses to a depth of $\gtrsim 1$ percent, which occurred in 90 percent of the examined synthetic light curves.
After both the selection process and the by-eye check for eclipses, only 0.3\% of the synthetic contact binary light curves remained.
We performed a similar by-eye check on the real targets selected after the cut in MMMR, removing 1455 of the 1501 targets.

We selected all targets with MMMR\,$> 0.8$ (using 0.8 rather than 1 to slightly alleviate the conservative nature of this cut).
We also applied a temperature cut to remove any star with $T_{\rm eff} > 6500\,K$ in order to facilitate easier measurement of RVs, where the $T_{\rm eff}$ estimate came from the \textit{TESS} Input Catalogue \citep[TIC;][]{Stassun2018,Stassun2019}, although in practice this removed relatively few targets.
\cmnt{This was originally done on the grounds that it's harder to measure the RVs of these, but it's an annoying inconsistency that I did this only for one selection method and not the other -- that was really just a bug that I would have changed if I'd realised.}
After these cuts and the by-eye eclipse check, 46 targets remained in the \mmmr\ sample of BH-LCs.

\section{Observations}
\label{sec:observations}

\begin{table}
\caption{A summary of the observations undertaken and archival data retrieved for this project. }
\begin{tabular}{lcc}
\hline
Data source & Date range & Num.~targets \\
\hline
\textit{Observations}\\
NTT & 2022 Jan 19--21 & 17\\
INT & 2022 Feb 14--16 & 19\\
INT & 2022 Jun 14--16 & 24\\
INT & 2024 Apr 26--30 & 1\\
\\
\textit{Archival data}\\
\textit{Gaia} SB1 & -- & 45 (8)\\
\textit{Gaia} SB2 & -- & 5 (1)\\
\textit{Gaia} $\Delta_{RV}$ & -- & 220 (71)\\
\\
\textit{Total} & -- & 250\\
\hline
\end{tabular}
\tablefoot{The number in parentheses refers to the number of targets in that category overlapping with targets in categories higher in the table.}
\label{tab:obsns}
\end{table}

\begin{table*}[]
\centering
\caption{Results for all targets for which observations were obtained, as is described in Sections~\ref{sec:observations}--\ref{sec:gaia}. }
\begin{tabular}{ccccccccccc}
\hline
TIC ID & Telescope & Epochs & \porb\ & $K$ & $f_M$ & $M_{\rm 2, min}$ & $q_{\rm min}$ & SB2 & $p_{\rm BH, 3}$ & $p_{\rm BH, 8}$\\
&&&[days] &  [km s$^{-1}$] & [$M_\odot$] & [$M_\odot$]\\
\hline
1947924 & \textit{Gaia} RVs & 29 & 1.95 & 36.7 & 0.010 & 0.33 & 0.17 & N & 0.02 & 0.01 \\
3103024 & \textit{Gaia} RVs & 11 & 2.06 & 35.2 & 0.009 & 0.34 & 0.17 & N & 0.02 & 0.01 \\
5769943 & NTT & 2 & 1.35 & $98.1 \pm 24.8$ & $0.132 \pm 0.127$ & $0.60 \pm 0.15$ & $0.47 \pm 0.12$ & N & 0.11 & 0.04 \\
7264195 & \textit{Gaia} RVs & 38 & 0.42 & 201.2 & 0.351 & 0.77 & 0.67 & N & 0.21 & 0.08 \\
13205508 & \textit{Gaia} RVs & 36 & 1.51 & 28.2 & 0.003 & 0.23 & 0.12 & N & 0.01 & 0.00 \\
18750830 & INT & 6 & 0.40 & $45.7 \pm 1.9$ & $0.004 \pm 0.001$ & $0.19 \pm 0.01$ & $0.14 \pm 0.01$ & Y & 0.01 & 0.00 \\
19124199 & NTT & 4 & 1.84 & $37.9 \pm 3.6$ & $0.010 \pm 0.003$ & $0.28 \pm 0.03$ & $0.19 \pm 0.02$ & N & 0.02 & 0.01 \\
21586667 & INT & 3 & 1.70 & $84.2 \pm 2.0$ & $0.105 \pm 0.008$ & $0.64 \pm 0.01$ & $0.40 \pm 0.01$ & N & 0.10 & 0.04 \\
24358586 & \textit{Gaia} RVs & 28 & 2.03 & 64.8 & 0.057 & 0.44 & 0.36 & N & 0.06 & 0.02 \\
26516212 & \textit{Gaia} SB1 & 25 & 2.63 & $19.8 \pm 3.2$ & $0.002 \pm 0.001$ & $0.19 \pm 0.03$ & $0.10 \pm 0.02$ & N & 0.01 & 0.00 \\
\vdots & \vdots & \vdots & \vdots & \vdots & \vdots & \vdots & \vdots & \vdots & \vdots  & \vdots \\
\hline
\end{tabular}
\tablefoot{The column $f_M$ contains the mass function.
This table contains 250 rows and is sorted by TIC ID.
The full table is available at the Centre de Données astronomiques de Strasbourg (CDS).
}
\label{tab:app-obs}
\end{table*}

\begin{table*}[]
\centering
\caption{Individual RV epochs obtained from our INT and NTT data, as is described in Sections~\ref{sec:observations}--\ref{sec:amplitudes}.}
\begin{tabular}{ccccc}
\hline
TIC ID & Telescope & BJD & RV [km s$^{-1}$] & $\sigma_{\rm RV}$ [km s$^{-1}$] \\
\hline
5769943 & NTT & 2459600.819035 & -59.76 & 1.57 \\
5769943 & NTT & 2459601.716446 & 29.74 & 1.19 \\
18750830 & INT & 2459625.455027 & 11.73 & 3.21 \\
18750830 & INT & 2459625.459147 & 47.13 & 3.03 \\
18750830 & INT & 2459625.462747 & 32.50 & 3.45 \\
18750830 & INT & 2459626.363180 & -26.07 & 3.29 \\
18750830 & INT & 2459626.366779 & -6.21 & 3.85 \\
18750830 & INT & 2459626.370390 & 0.13 & 3.94 \\
18750830 & INT & 2459626.374834 & -20.16 & 3.21 \\
18750830 & INT & 2459626.470489 & 23.06 & 3.27 \\
\vdots & \vdots & \vdots & \vdots & \vdots \\
\hline
\end{tabular}
\tablefoot{This table contains 478 rows and is sorted by TIC ID. The full table is available at the CDS.}
\label{tab:ind-rvs}
\end{table*}

We observed 60 of the targets selected in the previous section on the New Technology Telescope (NTT) and the Isaac Newton Telescope (INT). 
RVs for further targets were obtained from survey data, as is described in the following section.
A summary of the observations carried out, and survey data used, is given in Table~\ref{tab:obsns}.
A full list of all targets followed up or retrieved is given in Table~\ref{tab:app-obs}.

Targets selected by the {\tt mmmr} method were prioritised over the {\tt qmin} method, due to the higher purity of the former.
Priority was also given to brighter targets.
No priority was made on the value of $q_{\rm min}$, in order to aid the statistical considerations discussed in later sections (where we assume target selection is independent of the value of $q_{\rm min}$).

For targets observed on the NTT or INT, the observing strategy was to obtain three or more visits for each target (except for a minority of cases where observing constraints made only two visits possible).
A number of targets that were considered to be higher priority (those selected with the {\tt mmmr} method) were observed for between three and six visits.
No attempt was made to avoid repeat observations of SB2-type binaries or binaries with low amplitudes during a given observation run.
Each visit comprised three consecutive exposures.
The mid-exposure times of each exposure, as well as the RVs measured, are listed in Table~\ref{tab:ind-rvs}.

\begin{figure}
\centering
\includegraphics[width=0.99\hsize]{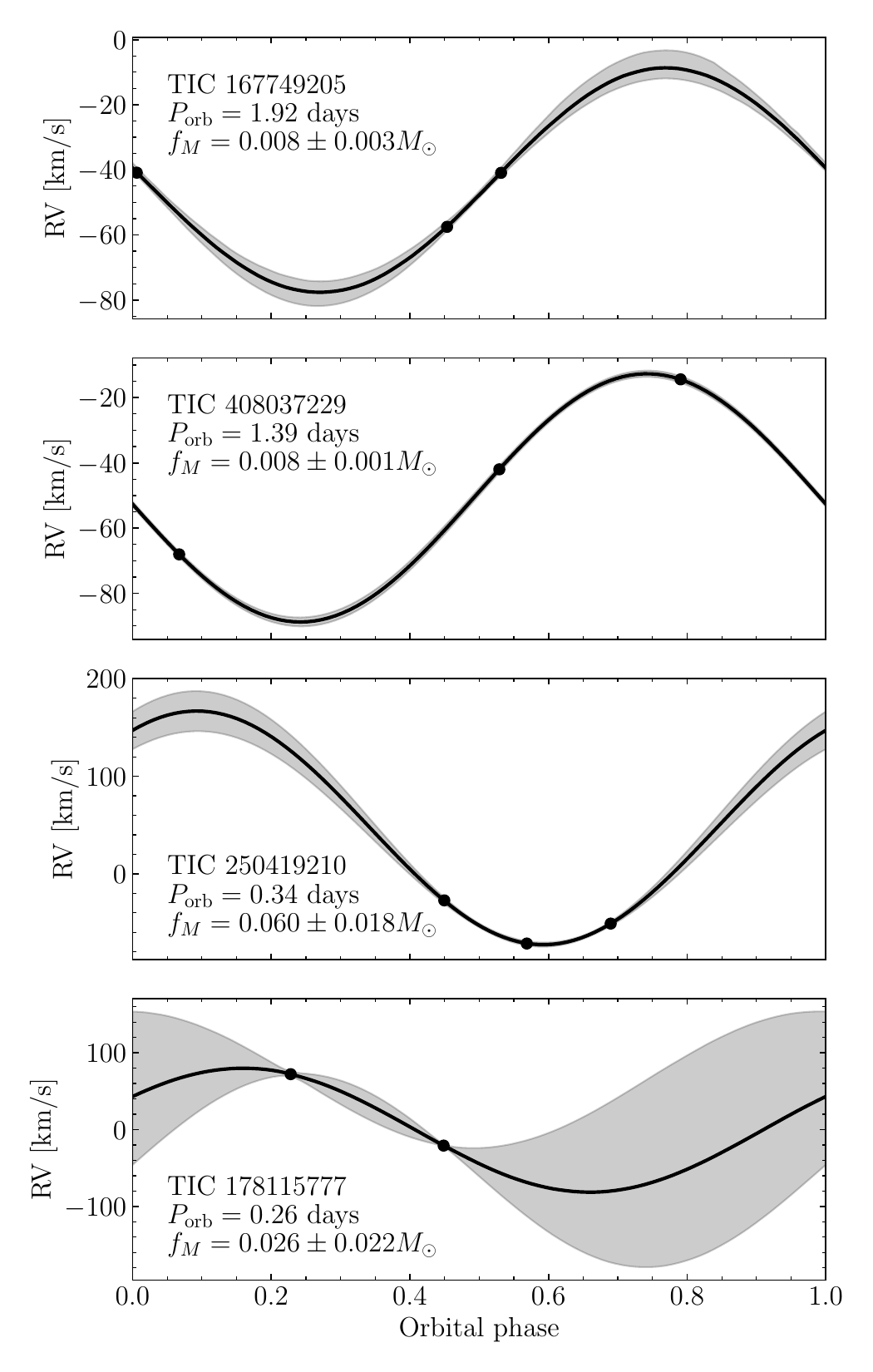}
\caption{
Orbital solutions for several example targets from our target list, all observed on the 2.5\,m INT.
TIC ID numbers are given in the figure panels, while $f_M$ is the spectroscopic mass function.
Black points are measured RV epochs, phase-folded on the photometric orbital period.
The shaded grey region shows the $1\sigma$ range of solutions found by the MCMC fitting process.
Three epochs is typically sufficient to find a precise orbital solution. 
Even in the example with only two epochs (last panel), the measurements are able to exclude $K$-amplitudes $\gtrsim 200$\,km s$^{-1}$.
}
\label{fig:example-rvs}
\end{figure}

Observations were carried out using the ESO Faint Object Spectrograph and Camera \citep[EFOSC;][]{Buzzoni1984}, a spectrograph mounted on the 3.5\,m NTT at La Silla Observatory in Chile. 
The Gr\#20 grism was used with a slit width of 0.5", yielding a wavelength range of 6000--7100\,\AA\ and a resolving power of approximately 3200.
A spectro-photometric standard, LTT\,1788\footnote{\url{cylammarco.github.io/SpectroscopicStandards}}, was observed on each night and used to flux-calibrate the spectra.

Further observations were carried out using the Intermediate Dispersion Spectrograph (IDS) instrument on the 2.5\,m INT at the Roque de los Muchachos Observatory on La Palma in Spain.
The R900V grating was used with a slit width of 1", yielding a wavelength range of 4400--5800\,\AA\ and a resolving power of 3200.
Flux standards SP0644+375 and SP1036+433 were observed in February and SP1436+277 in June and used to flux-calibrate the spectra.

Another target, TIC\,200292070, was identified as an interesting target at a late stage in the project.
Its {\tt rv\_ampltidue\_robust} measured by \textit{Gaia} (Section~\ref{sec:gaia-variance}) implied $q_{\rm min} \approx 2$.
We observed this target with the IDS/INT. 
The H1800V and R1200R gratings were used (due to set-up constraints unrelated to our observations) yielding resolving powers of 9400 and 6300, respectively.
Our measurements implied no RV variation; we have not been able to explain the \textit{Gaia} value.

Data were reduced using the {\tt \sc python} package {\tt \sc aspired} (Automated SpectroPhotometric Image REDuction)\footnote{\url{github.com/cylammarco/ASPIRED}} created by \citet{mlam2023} and \citet{mlam2023b}, following an optimal extraction routine \citep{Marsh1989}.
Each image was de-biased and flat-field corrected, and sky lines were subtracted with a polynomial fit.
Three consecutive spectra taken at each visit were combined to remove cosmic rays.
Wavelength calibration was performed using arc lamp spectra observed during the same visit, immediately following each set of exposures \citep{Veitch-Michaelis2020}.

\section{Measuring $K$-amplitudes}
\label{sec:amplitudes}

In order to measure the RV at each epoch, a cross-correlation was performed with a template spectrum.
The software package {\tt \sc sparta} \citep[SPectroscopic vARiabiliTy Analysis;][]{Shahaf2020}\footnote{\url{github.com/SPARTA-dev/SPARTA}} was used to perform the cross-correlation.
Spectra were continuum-normalised and the Balmer lines were masked so that the narrower metal lines dominated the cross-correlation, except in a select number of hotter stars for which the metal lines were too weak.

Template spectra were taken from the {\tt \sc phoenix} database \citep{Husser2013}.
Effective temperatures, surface gravities, and (where available) metallicities for our stars were taken from the TIC \citep{Stassun2019}, with a default value of 0 taken for metallicites that were not tabulated.
The closest available {\tt \sc phoenix}  template spectra to these properties were used for the cross-correlation.
Template spectra were broadened using a Gaussian kernel to match the resolution of the corresponding instrument.
We did not broaden the template to match the rotational velocity of the stars, even in cases where the rotational broadening was significant compared to the resolution (as was common among stars with \porb$ \lesssim 1$\,day due to tidal locking with the orbital period), on the grounds that excessive broadening can reduce the precision of the measured RVs.
The resulting RVs were measured to a precision of $\lesssim 1$\,km~s$^{-1}$.

Fifteen targets were identified as SB2-type, where visual inspection of the cross-correlation function showed clear evidence for two components in at least one epoch of observation.
We have identified those targets in a column of Table~\ref{tab:app-obs}.
We caution that these identifications are not necessarily complete, as identification depends on the epoch of observation as well as the clarity of the spectral lines from each star (which itself is dependent on the flux ratio and rotational broadening).

The measured RVs across all epochs were fit with an orbital model to determine the RV semi-amplitude ($K$).
All orbits were assumed to be circular given their short orbital periods \citep[e.g.][]{Zahn1977,Bashi2023}.
Priors were placed on the orbital period and phase, using the photometrically derived values from \citetalias{Green2023}.
Given these priors, we were able to perform the fit even for targets with only 2--3 epochs available, although for the small number of targets with only two epochs the resulting fit was poorly constrained.
An affine-invariant Monte-Carlo Markov Chain sampler \citep{Goodman2010,Foreman-Mackey2013} was used to explore the parameter space and determine uncertainties.
To account for the possibility that uncertainties on individual data may have been underestimated, we allowed the fit to include a common source of excess uncertainty $\sigma_{\rm exc}$, with a prior $\propto 1 / \sigma_{\rm exc}$, such that the effective uncertainty on each epochal RV $v_i$ is calculated from the formal uncertainty $\sigma_{v_i}$ by $\sqrt{\sigma_{v_i}^2 + \sigma_{\rm exc}^2}$.

Several examples are shown in Fig.~\ref{fig:example-rvs}.
The precision of the measured $K$-amplitude is driven primarily by the number of observed epochs (targets with only two epochs have much larger uncertainties), and secondarily by the phase coverage of the epochs.
We note that even large uncertainties on $K$ can be sufficient to exclude $K \gtrsim 200$\,km~s$^{-1}$, as in the example of TIC\,178115777\,shown here.
The temperature of the primary star impacted the precision of individual RV epochs, but we did not find it to be the dominant factor in the precision of $K$.

The resulting $K$-amplitudes are listed in Table~\ref{tab:app-obs}.
We also calculate lower limits on the companion mass and mass ratio.
These lower limits were found by calculating the spectroscopic binary mass function,
\begin{equation}
f_{\rm sp} = \frac{P_{\rm orb} K_1\,^3}{2 \pi G} = \frac{M_2^3 \sin^3 i}{(M_1 + M_2)^2},
\end{equation}
where $G$ is the gravitational constant, and then numerically solving for $M_2$ under the assumption that $\sin i = 1$. 

Assuming a geometric distribution of inclinations, it is possible to estimate the probability that a given companion would produce a given $K$-amplitude.
We also add columns showing the probability (implied by the measured mass function) that each target contains black holes of $M_2 = 3 M_\odot$ and $8 M_\odot$, calculated according to 
\begin{equation}
p_{\rm BH} = 1 - \cos i_{f_M} = 1 - \sqrt{1 - \frac{f_{M}^{2/3} (M_1 + M_2)^{4/3}}{M_2^2}},
\end{equation}
for mass function $f_M$, where $i_{f_M}$ is the orbital inclination that would be necessary to explain the measured $f_M$.

\section{\textit{Gaia} survey data}
\label{sec:gaia}

\begin{figure}
\centering
\includegraphics[width=0.99\hsize]{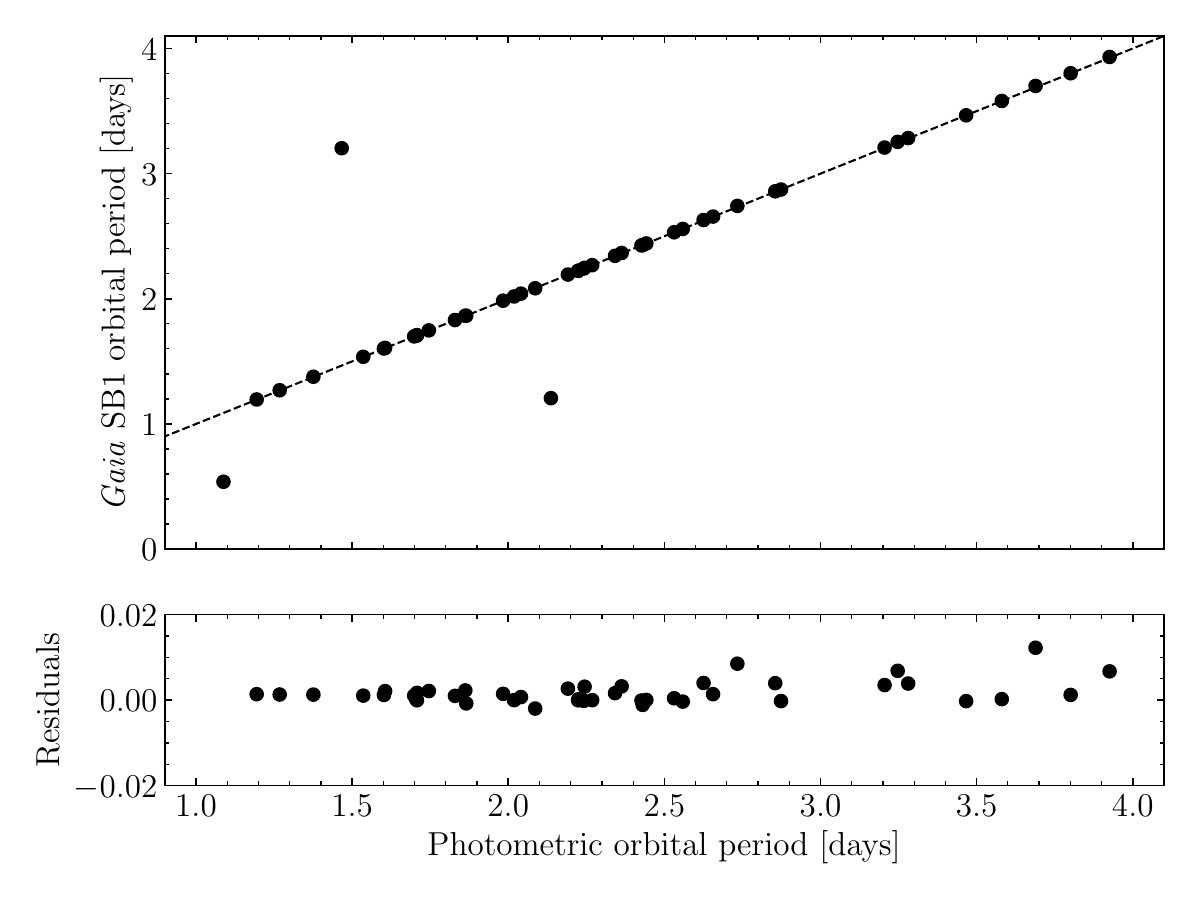}
\caption{
Verification of the \textit{Gaia} SB1 orbital solutions.
\citet{Bashi2022} previously noted that a number of \textit{Gaia} orbital solutions at short periods are unreliable due to aliasing issues.
Here we show that, of our photometrically selected targets that have \textit{Gaia} SB1 solutions, all but three are in agreement to within $<1$\%.
}
\label{fig:periods-compare}
\end{figure}

\begin{figure}
\centering
\includegraphics[width=0.99\hsize]{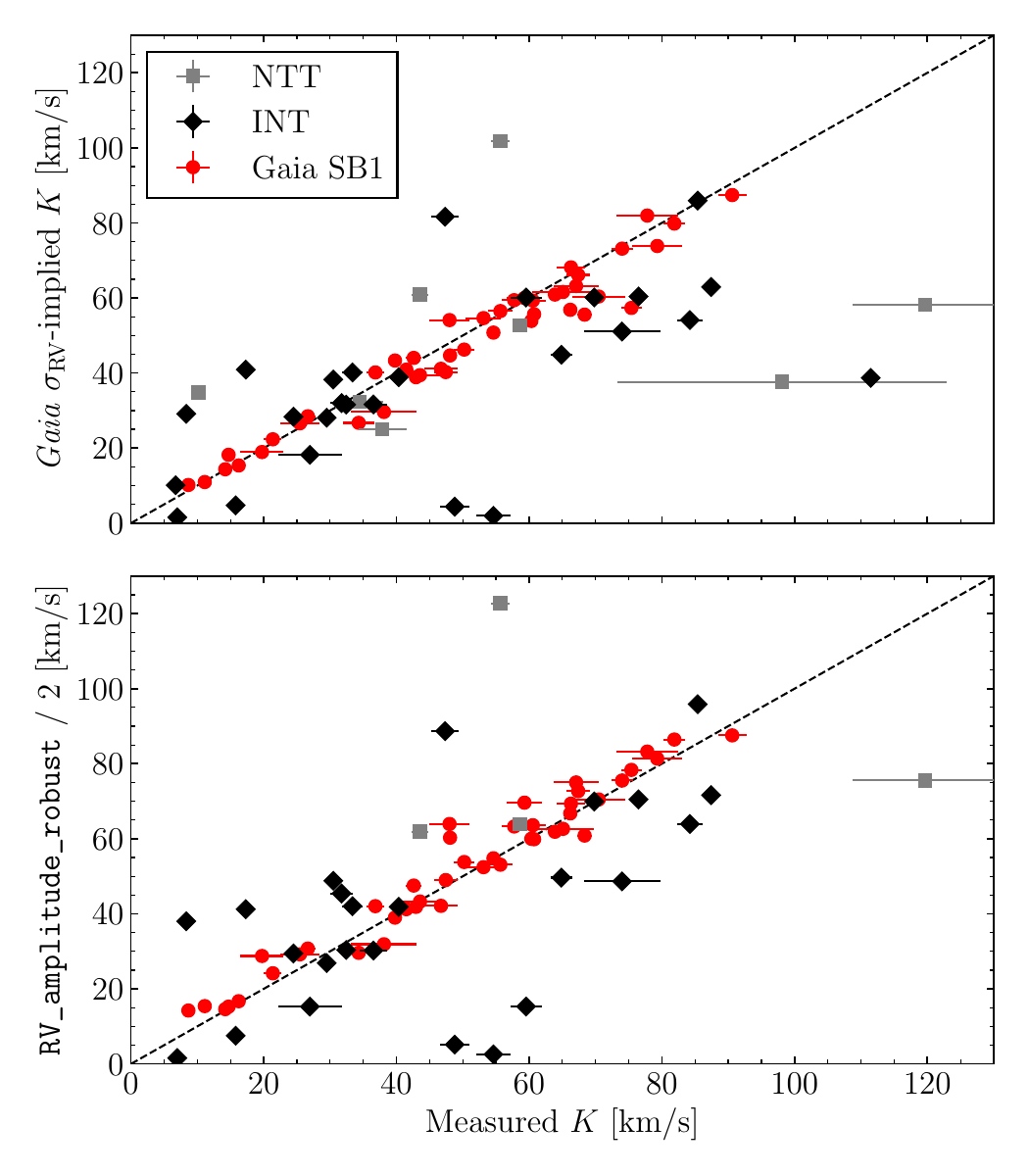}
\caption{
The $K$-amplitudes derived via two methods from \textit{Gaia} data, against those measured from orbital solutions, for all targets with both measurements.
There is generally reasonable agreement, with some outlying points, as is discussed in the text.
}
\label{fig:rv-comparison}
\end{figure}

\subsection{Spectroscopic orbital solutions}

To complement our observations, we retrieved data from the \textit{Gaia} Radial Velocity Spectrometer \citep[RVS; Data Release 3;][]{GaiaCollaboration2016,GaiaCollaboration2022}.
Although epochal spectroscopic data are not yet available for most targets, there are orbital solutions for a number of binary systems \citep{Halbwachs2022}.
The number of SB1 and SB2 orbital solutions available for our targets is summarised in Table~\ref{tab:obsns}.

\citet{Bashi2022} have noted that the \textit{Gaia} SB1 catalogue contains a number of spurious orbital solutions, especially at short orbital periods where aliasing issues are common.
Although they propose a recommended `clean' subset, it contains relatively few systems in our period range of interest (fewer than 1000 of their `clean' systems have $P_{\rm orb} < 3$\,days, and only 25 have $P_{\rm orb} < 1$\,day).
The `clean score' given by \citet{Bashi2022} for our targets is distributed close to uniformly between 0.1 and 0.8, and only eight of the 48 targets that have \textit{Gaia} SB1 solutions have clean scores above the recommended cut-off of 0.587.
However, given that the main contributor to unreliable \textit{Gaia} SB1 solutions is aliasing, we consider the independent photometric confirmation of the orbital period to be a confirmation of the orbital solution, and keep all solutions for which the \textit{Gaia} orbital period is consistent with the photometrically determined orbital period.
This was the case for 45 of the 48 targets from our target list with \textit{Gaia} SB1 solutions, as plotted in Fig.~\ref{fig:periods-compare}.
Of the nine \textit{Gaia} SB1 targets for which we have independently obtained velocity semi-amplitudes (after excluding targets for which the spectroscopic orbital period is not consistent with our photometric period), five have amplitudes that are consistent within $< 3.5 \sigma$, while the other four are significant outliers. 
All are within several tens of km\,s$^{-1}$, similar to the scatter in Fig.~\ref{fig:rv-comparison} (next section), which is sufficient precision to identify the expected RV of $\gtrsim 100$ km\,s$^{-1}$ that would be induced by a black hole companion.

\subsection{RV variance measured by \textit{Gaia}}
\label{sec:gaia-variance}

Information from the \textit{Gaia} DR3 catalogue can be used to derive an estimate of the RV variability for vast numbers of stars.
There are two approaches that can be used here.
Firstly, the \textit{Gaia} tables include an estimate of the peak-to-peak RV amplitude, {\tt RV\_amplitude\_robust}.
Secondly, the tabulated mean RV in the \textit{Gaia} database has an associated uncertainty, which is calculated from the standard deviation of individual RV measurements, $\sigma_{\rm RV}$, according to\footnote{Taken from the \textit{Gaia} DR3 documentation, \url{https://gea.esac.esa.int/archive/documentation/GDR3/Gaia_archive/chap_datamodel/sec_dm_main_source_catalogue/ssec_dm_gaia_source.html}}
\begin{equation}
{\tt dr3\_radial\_velocity\_error} = \sqrt{\frac{\pi}{2}\frac{\sigma_{\rm RV}^2}{n_{\rm RV}} + 0.11^2},
\end{equation}
where $n_{\rm RV}$ is the number of epochs.
This can be inverted to find the RV variance $\sigma_{\rm RV}$, and the amplitude of a circular orbit can be estimated as $K = \sqrt{2} \sigma_{\rm RV}$.
Both methods can only be applied to targets with ${\tt RV\_method\_used} = 1$.

We used these two methods to estimate $K$ for all targets that have tabulated \textit{Gaia} RVs with uncertainties based on more than five epochs.
In Fig.~\ref{fig:rv-comparison}, we compare the $K$-amplitudes measured from orbital solutions to those estimated in these two ways.
There is generally reasonable agreement between the solution amplitudes and the estimated amplitudes, although there are a number of outliers.
There were two outlying systems for which we measured clear RV variability that was not reported by \textit{Gaia}; we have not found an explanation for this.
We suggest that several outliers (especially those above the trend) may be unrecognised SB2s, for which line blending differently affected the redder \textit{Gaia} RVS spectra and our bluer spectra.
While the two \textit{Gaia} approaches have very similar results, we favour the $\sigma_{\rm RV}$ method, which produces slightly fewer outliers (21/78 compared to 26/78 of the targets in Fig.~\ref{fig:rv-comparison}).

Not all of our targets have tabulated RVs in \textit{Gaia}.
The tabulated RVs may be missing for a number of possible reasons: colour and magnitude selection effects; if the amplitude is particularly large; or if the target appeared to have multiple spectral components.
It is therefore difficult to attribute a unique explanation for any one missing value.
The $K$-amplitudes retrieved from \textit{Gaia} orbital solutions or estimated from $\sigma_{\rm RV}$ are tabulated in Table~\ref{tab:app-obs}.

\subsection{Summary of $K$-amplitudes}

\begin{figure}
\centering
\includegraphics[width=0.9\hsize]{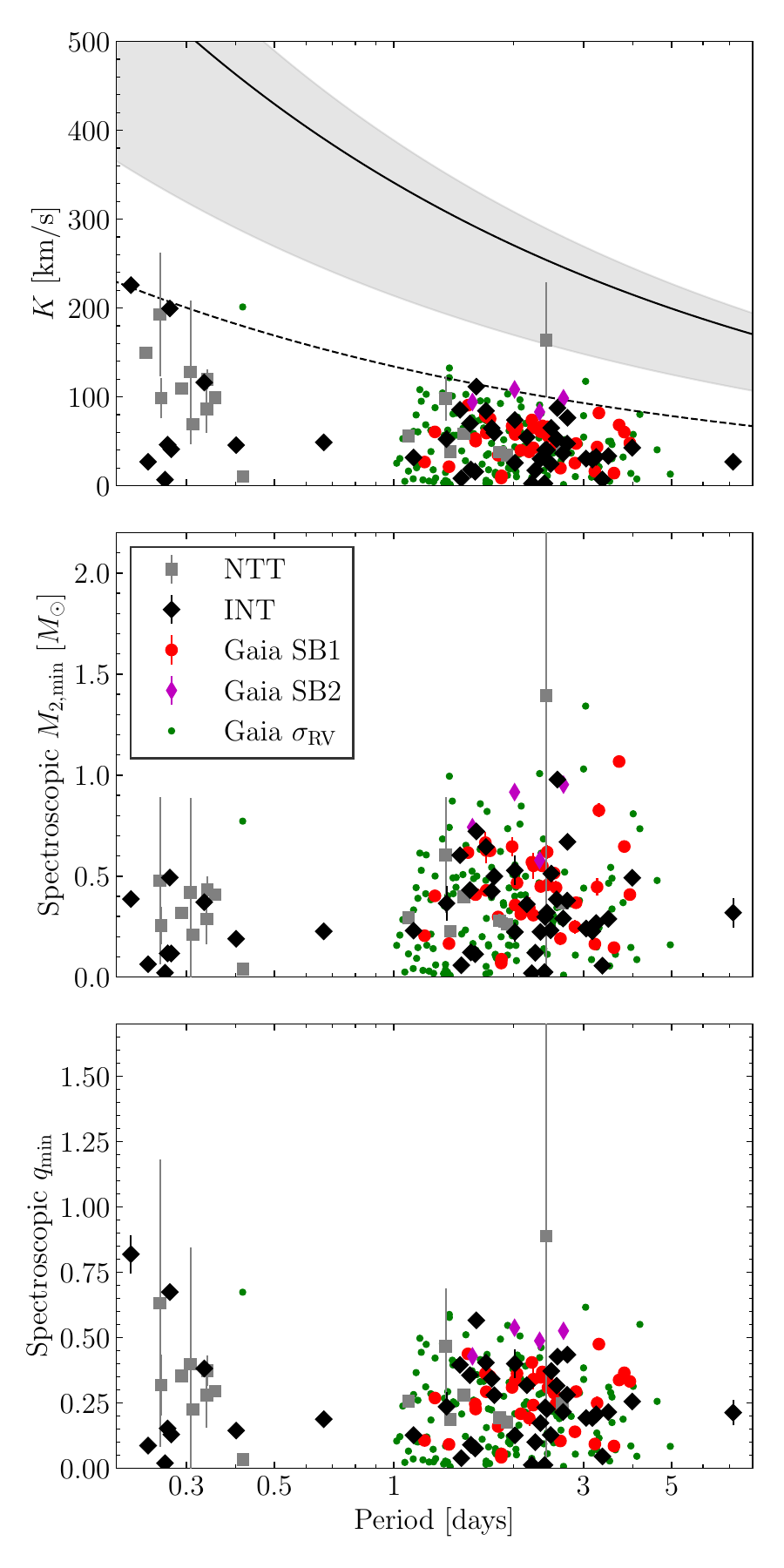}
\caption{Measured $K$-amplitudes \textit{(top)} and the implied lower limits on $M_2$ \textit{(middle)} and $q$ \textit{(bottom)} for our observed targets. 
Also plotted in the top panel are the expected $K$-amplitudes for a $1 M_\odot$ star with an $8 M_\odot$ companion at the median orbital inclination (solid black line) and the central 64 percent range (shaded region) for a geometric distribution of inclinations; and the maximum expected amplitude for an equal-mass binary (dashed line).
The deficiency of targets in the period range 0.5--1\,days is due to the different selection methods applied at shorter and longer periods, as is described in the text.
None of the systems followed up has spectroscopic $M_{\rm 2, min} > 3\,M_\odot$ or $q_{\rm min} > 1$, and in most cases a typical-mass black hole can be ruled out at the $2\,\sigma$ level.
}
\label{fig:k-amplitudes}
\end{figure}

Using the methods described in the previous sections, we have measured or estimated $K$-amplitudes for 250 of our 457 BH-LC targets.
The $K$-amplitudes and derived companion mass limits are plotted in Fig.~\ref{fig:k-amplitudes}.
Overall, none of our candidates followed up so far remains a promising BH-LC candidate.
Even our largest $K$-amplitudes are difficult to explain as originating from a high-mass companion, unless the inclination is unusually face-on.
The presence of an $8\,M_\odot$ ($3\,M_\odot$) companion is excluded at a $\gtrsim 2 \sigma$ (1.5\,$\sigma$) level for more than 90 percent of our targets, and is excluded at the $1 \sigma$ level in all cases.
Having all but ruled out BH-LC binaries among these 250 candidates, we can estimate an upper limit on the space density of short-period BH-LCs, as is described in the following section.

\section{The fraction of stars that are short-period BH-LCs}
\label{sec:upper-lim}

In this section, we describe the process of converting this non-detection to an upper limit on the BH-LC space density.
We first overview the conversion in a formal manner, before describing the numerical integration performed to find the selection efficiency.

\subsection{Formal description}

\begin{figure}
\centering
\includegraphics[width=\hsize]{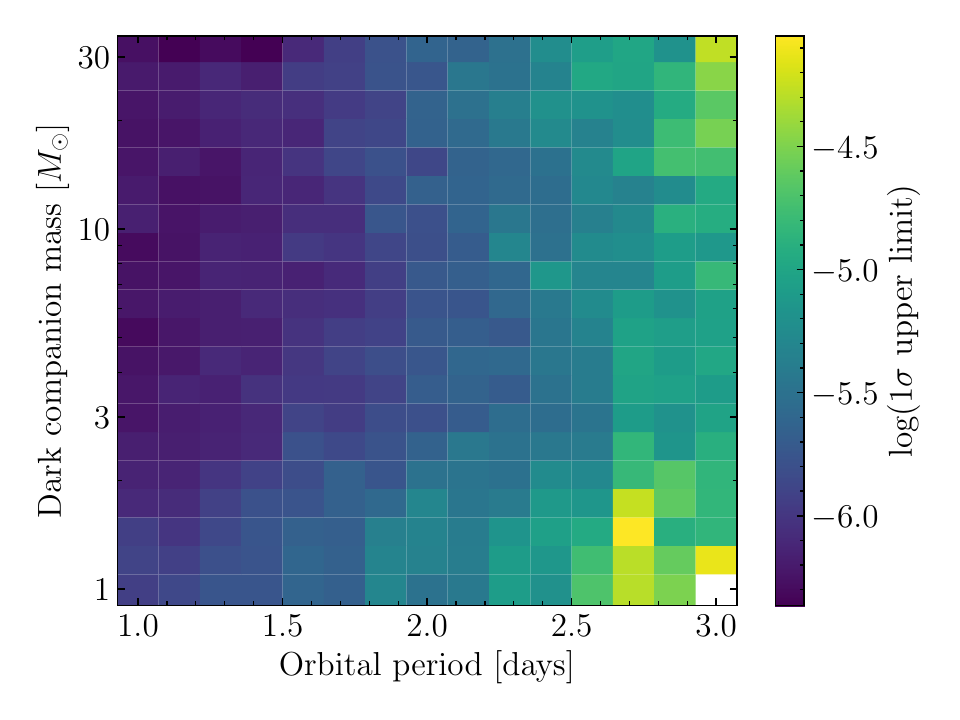}
\caption{
Two-dimensional upper limits on the frequency of black hole companions to solar-type stars as a function of orbital period and black hole mass, $f_{\rm BH} (M_2, P_{\rm orb})$.
The existence of orbital periods close to 1\,day is more tightly constrained than the existence of periods close to 3\,days.
Dark companions with masses $M_2 < 3 M_\odot$ are less tightly constrained than more massive companions, but above this limit the dependence on companion mass is weak.
\cmnt{Is it true that high-mass bhs are harder to find at fixed period?}
}
\label{fig:upperlim-2d}
\end{figure}

\begin{figure*}
\resizebox{\hsize}{!}
{\includegraphics{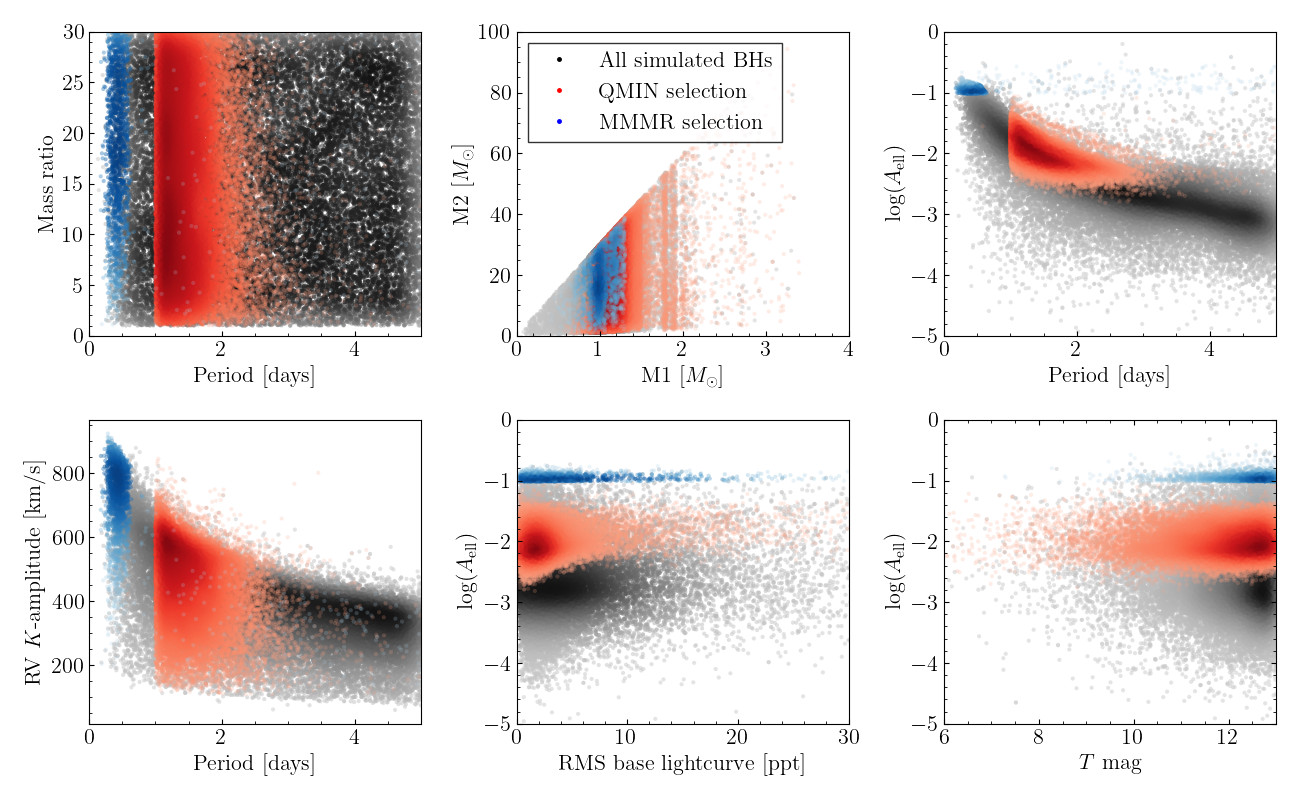}}
\caption{Properties of the simulated BH-LC population, with coloured highlights showing the subsets of systems that were selected using the \mmmr\ and \qmin\ methods. The panels show the input distributions of \porb, $q$, $M_1$ and $M_2$, the distributions of the measured properties \aell\ and $K$, and the behaviour of \aell\ as a function of $T$-band apparent magnitude and the RMS of the base light curve. The latter property includes shot noise which depends on SNR (and hence is related to $T$-band magnitude), but also includes additional scatter due to \textit{TESS} instrument systematics or intrinsic stellar variability.
}
\label{fig:simulation-summary}
\end{figure*}

\begin{figure}
\centering
\includegraphics[width=\hsize]{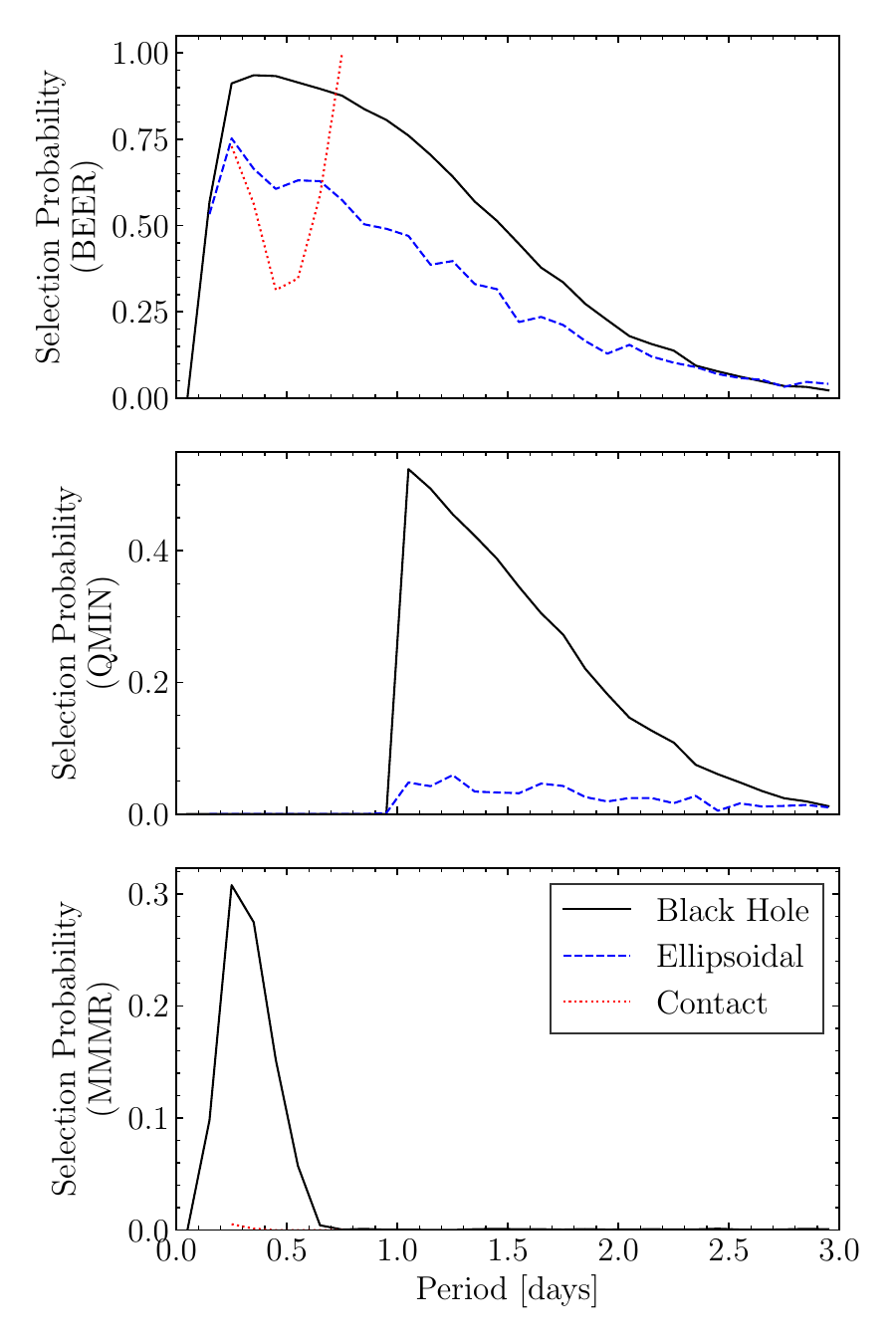}
\caption{
Probability for simulated binary systems of various types to be accepted into the \beer\ sample (top), \qmin\ sample (middle), and \mmmr\ sample (bottom), as a function of period. 
Other variables ($\cos i$, $M_1$, and $M_2$) have been marginalised over.
}
\label{fig:selection-function-beer}
\end{figure}

We assume that, for a given target, the primary star (or the single star if it is not in a binary system) can be described by a vector $\Vec{Y} = (M_1, R_1, m_T)$, where $m_T$ is the \textit{TESS}-band apparent magnitude.
A black hole companion will have mass $M_2$, and the system orbit will be described by $\cos i$ and \porb. 
Given the short periods, we assume all orbits to be circular \citep[e.g.][]{Bashi2023}.

We aim to define a model of a BH-LC population and predict how many BH-LCs would be detected as such under that model.
Because there are very few data with which to constrain the BH-LC population, we assume the simplest possible model: a constant fraction $f_{\rm BH}$ of luminous stars has a black hole companion, independent of the properties of that luminous star.

The directly observable properties of the binary are $m_T$, \porb, and \aell, where $A_{\rm ell} = A_{\rm ell} (M_1, M_2, R_1, P_{\rm orb}, \cos i)$ according to Equation~\ref{eq:ell2}.
We define a selection function, which determines the probability that a binary with a given set of observable properties will be selected into our BH-LC candidate list (\qmin\ or \mmmr), as
\begin{equation}
p(\text{select}\,|\, A_{\rm ell}, P_{\rm orb}, m_T) = S_{\rm obs} (A_{\rm ell}, P_{\rm orb}, m_T),
\end{equation}
which, under the assumptions detailed above, can be rewritten in terms of the binary's physical properties,
\begin{equation}
p(\text{select}\,|\, \Vec{Y}, M_2, P_{\rm orb}, \cos i) = S_{\rm phys}(\Vec{Y}, M_2, P_{\rm orb}, \cos i).
\label{eq:pselect}
\end{equation}
The estimation of $S_{\rm phys}$ was carried out numerically using a set of injection-recovery tests, as is described in Section~\ref{sec:simulations}.

We can marginalise over all values of $\cos i$ (which is uniform for a geometric distribution of orbital inclinations) to find the average probability that a given black hole with $M_2$ and \porb\ would be detected around a given host star:
\begin{equation}
p(\text{select}\,|\, \Vec{Y},   M_2,  P_{\rm orb}) = \int d \cos i \, S_{\rm phys}  (\Vec{Y}, M_2, P_{\rm orb}, \cos i).
\end{equation}
If we then assume that a fraction $f_{\rm BH} (M_2,P_{\rm orb})$ of stars have black hole companions with $M_2$ and $P_{\rm orb}$, we would expect the number of detached BH-LCs with $M_2$ and $P_{\rm orb}$ selected to our list of candidates to be
\begin{align}
N_{\rm BH, selected} (M_2,P_{\rm orb}) &= f_{\rm BH} (M_2,P_{\rm orb}) \sum^{N_{\rm input}}_{j=1} p(\text{select}\,|\, \Vec{Y}_j, M_2,  P_{\rm orb})\\
&= f_{\rm BH} (M_2,P_{\rm orb}) \, N_{\rm input} \Bar{S}(M_2,  P_{\rm orb}),
\end{align}
where $N_{\rm input}$ is the number of stars in the \inp\ sample, $j = \{ 1 ... N_{\rm input} \}$ represents the set of all stars and $\Bar{S}(M_2,  P_{\rm orb})$ represents the average value of $p(\text{select}\,|\, \Vec{Y}_j, M_2,  P_{\rm orb})$ across all stars in the \inp\ sample and has a value between 0 and 1.

In our case, we were not able to obtain meaningful follow-up observations for all BH-LC candidates that were selected.
We can therefore introduce another fraction $f_{\rm observed}$, such that the number of black holes with a given $M_2$ and $P_{\rm orb}$ that were finally observed is 
\begin{align}
N_{\rm BH, observed} (M_2, P_{\rm orb}) = &f_{\rm observed} N_{\rm BH, selected} \\
= &f_{\rm observed} \, f_{\rm BH}  \, (M_2, P_{\rm orb})  N_{\rm input} \Bar{S}(M_2,  P_{\rm orb}).
\label{eq:fbh}
\end{align}
We can then find $f_{\rm BH}$ directly by inverting Eq.~\ref{eq:fbh}.
In order to find the 1, 2, and 3$\sigma$ upper limits on $f_{\rm BH}$, we can put values of $N_{\rm BH, observed} = 1.0$, 3.9, and 8.7 \citep[values that disagree with zero at the 1, 2, and 3$\sigma$ levels according to a Wilson score interval;][]{Wilson1927} into Equation~\ref{eq:fbh}.

This leads to a two-dimensional distribution of upper limits, based on our {\tt qmin} sample, shown in Fig.~\ref{fig:upperlim-2d}. 
Worthy of note is that our upper limit is nearly independent of the actual black hole mass for masses $M_2 \gtrsim 3 M_\odot$.
The upper limit is strongest at periods $\sim 1$\,day, where we can say that a fraction $f_{\rm BH} (M_2 > 3 M_\odot, P_{\rm orb} = 1\,{\rm day}) < 10^{-6}$ of stars can have black hole companions.

In order to arrive at an overall \fbh\ independent of black hole parameters, we must marginalise over some assumed prior distributions of $p(M_2)$ and $p(P_{\rm orb})$.
The overall probability of detecting a black hole around any given star is then
\begin{equation}
p(\text{select}\,|\, \Vec{Y}) = \int d \cos i \int d P_{\rm orb} \, p(P_{\rm orb}) \int d M_2 \, p(M_2)  \, S_{\rm phys}.
\label{eq:integrals}
\end{equation}
Summing over each host star, we then expect 
\begin{equation}
N_{\rm BH, selected} = f_{\rm BH} \sum^{N_{\rm input}}_{j=1} p(\text{select}\,|\, \Vec{Y}_j) = f_{\rm BH} \, \Bar{S}' N_{\rm input},
\label{eq:sum}
\end{equation}
and
\begin{equation}
N_{\rm BH, observed} = f_{\rm observed} N_{\rm BH, selected} = f_{\rm observed} f_{\rm BH} \, \Bar{S}' N_{\rm input},
\label{eq:fbh2}
\end{equation}
where $\Bar{S}'$ represents the average value of $p(\text{select}\,|\, \Vec{Y}_j)$ across all stars in the \inp\ sample. 
This $\Bar{S}'$ implicitly contains the integrals in Equation~\ref{eq:integrals} and has a value between 0 and 1.

The value of \fbh\ can be found by inverting Equation~\ref{eq:fbh2} and inserting the Wilson score interval values, as before.
The values of \sbar\ were calculated numerically using the method described in the following section, and are listed in Table~\ref{tab:sbar}.
Putting \sbar$_{\rm Overall}$ into Equation~\ref{eq:fbh}, we can derive $f_{\rm BH} < 2.4 \times 10^{-6}$ ($1\,\sigma$), $9.5 \times 10^{-6}$ ($2\,\sigma$), and $21 \times 10^{-6}$ ($3\,\sigma$).

\subsection{Simulated population}
\label{sec:simulations}

To calculate \fbh, as was laid out in the previous section, it is necessary to estimate the efficiencies of our selection methods as a function of the binary physical parameters, $S_{\rm phys}(\Vec{Y}, M_2, P_{\rm orb}, \cos i)$.
We estimated this by simulating a population of BH-LCs and performing injection-recovery tests using their synthetic light curves.
In fact, a shortcut is possible: 
if the simulated population is drawn from the desired distributions $p(M_2)$ and $p(P_{\rm orb})$, and the distribution of $\Vec{Y}_j$ is the same as the \inp\ sample, then we can simply use the fraction of simulated binaries that are selected as an estimate of the average selection efficiency $\Bar{S}'$. 
Effectively this is a numerical calculation of the integrals in Equations~\ref{eq:integrals}--\ref{eq:sum}.

We have three samples of candidates that are drawn from the \inp\ sample: the \beer, \qmin, and \mmmr\ samples.
Each of these has its own value of $\Bar{S}'$, which we refer to as $\Bar{S}'_{\rm BEER}$, $\Bar{S}'_{\rm QMIN}$, and $\Bar{S}'_{\rm MMMR}$.


The simulated BH-LC population was generated following a similar methodology to the injection-recovery tests in \citetalias{Green2023}.
First, a primary star was chosen at random from among the \inp\ sample.
The primary star temperature was taken from the TIC \citep{Stassun2019}, and $M_1$ and $R_1$ were estimated by interpolation of the tables of \citet{Pecaut2013}, under the assumption that the primary star is on the main sequence.
$M_2$ and \porb\ were drawn from assumed distributions $p(M_2)$ and $p(P_{\rm orb})$, see below, while $\cos i$ was drawn from a uniform distribution between 0 and 1. 
Using the drawn parameters, a synthetic ellipsoidal binary light curve was generated, with \aell\ calculated according to Equation~\ref{eq:ell2} and amplitudes at other harmonics calculated according to the description in \citetalias{Green2023}.
The synthetic light curve was added to the base \textit{TESS} light curve of the selected primary star, which ensures that the noise profile, \textit{TESS} systematics, and any underlying stellar variability, will all be representative of a star at that temperature and magnitude.
This process was repeated until the desired population size was reached.
Any simulated binaries in which the primary star would overfill its Roche lobe were discarded.

Two unknowns are the distributions $p(M_2)$ and $p(P_{\rm orb})$. 
We assumed a uniform distribution between 0 and 3 days for \porb.
Systems with periods shorter than $\approx 0.3$\,days are typically removed later due to being Roche-lobe filling, but this was individually tested in each case. 
For $q$ we assumed a uniform distribution between 1 and 30, with $M_2$ then calculated from $q$ and $M_1$.
For the calculation of selection probabilities we filtered out all simulated binaries with $M_2 < 3$, as masses below this limit are not expected for black holes.
We note that the selection probability does not strongly depend on $q$ for $q > 3$, as is shown in Fig.~\ref{fig:upperlim-2d}.\footnote{For example, when the range of $M_2$ is limited only to masses 5--10\,$M_\odot$, $\bar{S}'_{\rm Overall}$ changes from 0.180 to 0.171, a negligible change compared to other uncertainties.}
Several summary plots of the simulation are shown in Fig.~\ref{fig:simulation-summary}.

Once the simulated populations and synthetic light curves were produced, we performed injection-recovery tests by feeding each light curve through our candidate selection process.
First, each light curve was analysed using the {\sc beer} algorithm, as is described in \citetalias{Green2023}, and either selected or not as a \beer\ candidate.
The fraction of selected systems gives an estimate of $\Bar{S}'_{\rm BEER}$.
After this, targets were selected or not into the \qmin\ and \mmmr\ samples on the basis of the \aell\ measured by the {\sc beer} algorithm, allowing an estimate of $\Bar{S}'_{\rm QMIN}$ and $\Bar{S}'_{\rm MMMR}$.
In the case of the \qmin\ method, the values of $M_1$ and $R_1$ used in this calculation were shifted from the `true' input values of the synthetic binary system by an amount drawn from a Gaussian uncertainty profile of the corresponding width, in order to approximate the uncertainty on $M_1$ and $R_1$ in the true binaries.

In Fig.~\ref{fig:selection-function-beer}, we show the selection functions of each method for the simulated population as a function of period.
We also performed similar injection-recovery tests for simulated populations of detached MS-MS and contact binaries that were originally created for \citetalias{Green2023} (see that paper for a full description of those simulations).
These are also included in Fig.~\ref{fig:selection-function-beer}.
As may be expected from the discussion in previous sections, the \qmin\ method is most sensitive to periods $\approx 1$\,day (with a sharp cut-off due to the period cut applied in that method), while the \mmmr\ method is most sensitive to periods $\lesssim 0.5$\,days.

\begin{table}
\caption{Selection efficiencies for black holes in our selection methods, estimated from injection-recovery tests.}
\begin{tabular}{lr}
\hline
$\bar{S}'$ & Value\\
\hline
$\bar{S}'_{\rm BEER}$ & 0.42\\
$\bar{S}'_{\rm MMMR}$ & 0.010\\
$\bar{S}'_{\rm MMMR}$ ($P_{\rm orb} < 0.5$) & 0.040\\
$\bar{S}'_{\rm QMIN}$ & 0.17\\
$\bar{S}'_{\rm QMIN}$ ($P_{\rm orb} > 1$) & 0.31\\
$\bar{S}'_{\rm Overall}$ & 0.18\\
\hline
\end{tabular}
\label{tab:sbar}
\end{table}





\section{Discussion}
\label{sec:discussion}

\subsection{Comparison to theoretical predictions}

\begin{figure}
\centering
\includegraphics[width=\hsize]{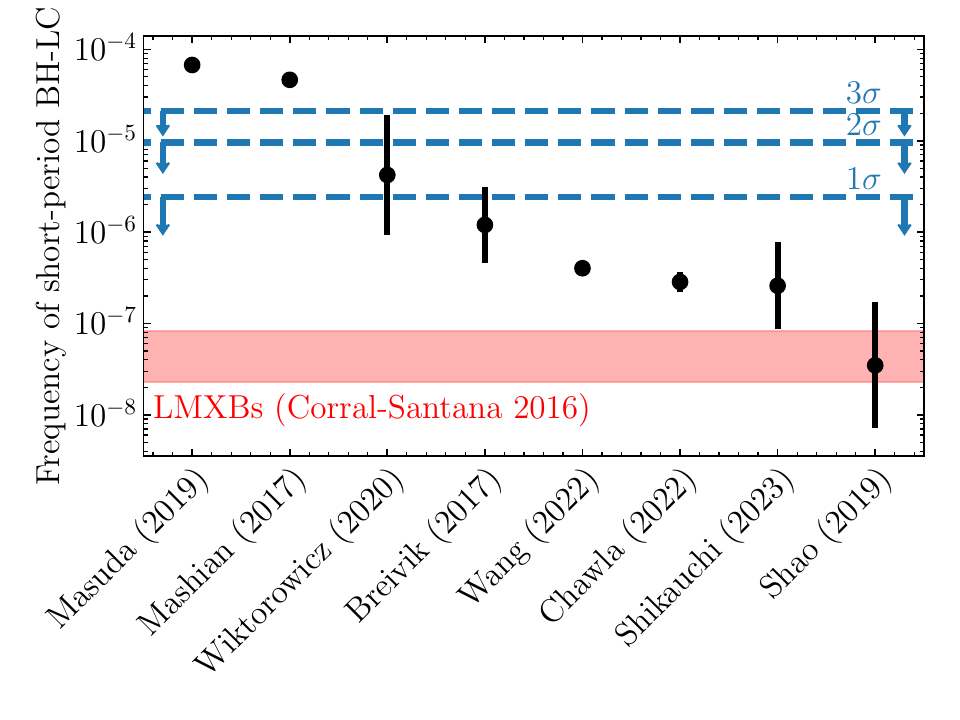}
\caption{Upper limits on the frequency of short-period BH companions to solar-type stars derived from this work (horizontal dashed lines), compared to various theoretical predictions from the literature, with error bars showing the range of theoretical predictions when multiple values were reported within one work. 
Our upper limits can reject the more optimistic predictions. Also shown (red band) is an observational estimate of the overall LMXB fraction in the Galaxy.
}
\label{fig:compare-literature}
\end{figure}

A number of efforts have been made in recent years to predict the occurrence rate and detectability of BH-LCs in the Galaxy.
Most of these efforts have specifically aimed to predict the BH-LC samples to be detected by \textit{Gaia}, but in some cases the predictions can be generalised and compared to our \textit{TESS}-based sample results.

An early model by \citet{Mashian2017} made a number of simplifying assumptions: they neglected binary interactions, assumed a log-uniform distribution of orbital separations, and assumed that all stars with mass $ > 20 M_\odot$ will become BHs with no natal kick.
\citet{Breivik2017} moved beyond these assumptions with a binary population synthesis model that incorporated binary interactions and several possible distributions of natal kicks.
Later models have investigated the impacts of extinction \citep{Yamaguchi2018}, different prescriptions for the common envelope phase of binary interaction \citep{Yalinewich2018,Shao2019}, star formation histories and chemical evolution \citep{Wiktorowicz2020}, delayed and rapid supernova scenarios \citep{Chawla2022,Shikauchi2022}, and Galactic location with its associated age and metallicity dependencies \citep{Chawla2022,Wang2022,Shikauchi2023}.

Additional work by \citet{Masuda2019} made a direct prediction for the detection of ellipsoidal binary systems in \textit{TESS}.
They investigated two scenarios: one in which the BH-LC population resembles the field MS-MS binary population \citep[similar to][]{Mashian2017}; and one in which binary mass transfer interactions were incorporated. 
Interestingly, they found that the rate of short-period BH-LCs was similar between the two scenarios because binaries that merged in the scenario that allowed interactions were replaced by binaries with initially longer periods that had moved inwards during the common envelope phase.
Neither scenario included any binary disruption due to natal kicks.
They predicted that 400--450 BH-LCs would be detectable as ellipsoidal binaries in the \textit{TESS} dataset, assuming a somewhat fainter magnitude limit ($T < 15$) than ours ($T < 13.5$).


Also worth noting---though not immediately relevant to this work---are \citet{Andrews2019}, who used mock \textit{Gaia} data to investigate selection effects in more detail; \citet{Shikauchi2020}, who modelled BH-LC formation in dense stellar environments; and \citet{Janssens2022,Janssens2023}, who performed population synthesis models focussing on BH companions to high-mass stars.

In most of the works referenced above, the predictions are not presented in a form that can be directly compared to our upper limit, but must be converted to some common metric.
We adopt as our metric `frequency of black hole companions to MS stars with \porb\,$< 3$ days' (henceforth $f_{\textrm{BH}; P<3}$) and convert the predictions of those papers to this metric, although doing so requires a number of simplifying assumptions.
For papers that quoted a total number of BH-LCs in the Galaxy, we take this number and divide it by the approximate number of solar-type stars in the Galaxy ($2 \times 10^{10}$).
Information about Galactic location and host stellar type is therefore lost, even from models that originally considered it.
For \citet{Wang2022}, we take their tabulated number of BH-LCs that are in the thin disc (since the vast majority of our targets are thin disc sources). 
A further assumption must be made about the orbital period distribution.
In many of the works cited above \citep[e.g.][]{Shao2019,Shikauchi2023} the orbital period distribution of the population is close to log-uniform.
Therefore, for papers that quote the numbers of BH-LCs in the Galaxy across all orbital periods, we assume a log-uniform distribution between periods of 0.3\,days and $10^5$\,years and use this to determine how many are within our period range of \porb\,$< 3$\,days.
For papers that quote numbers within limited period ranges \citep{Shao2019,Chawla2022,Shikauchi2023}, we extrapolate to our period range from the shortest-period range quoted in their papers using the same log-uniform assumption.
Given these caveats, the conversion to $f_{\textrm{BH}; P<3}$ is rather approximate but should suffice for an order-of-magnitude comparison, which is valuable given that the predictions span multiple orders of magnitude.
The resulting values of $f_{\textrm{BH}; P<3}$ are listed in Table~\ref{tab:literature}. 
In the subsequent discussion we denote the models that predict larger numbers of detectable BH-LC binaries as `optimistic', and those that predict smaller numbers as `pessimistic'.


In Fig.~\ref{fig:compare-literature}, we compare the estimated values of $f_{\textrm{BH}; P<3}$ from Table~\ref{tab:literature} with our upper limits.
It can be seen that the most optimistic models are strongly ruled out by our observations.
Intermediate predictions, such as \citet{Breivik2017} and \citet{Wiktorowicz2020}, are somewhat challenged at the level of 1-2$\,\sigma$.
More pessimistic predictions, including all those more recent than 2020, are still an order of magnitude lower than what can be tested with the current data.
The upper limit we have derived here is similar to the one on longer-period BH-LC binaries, estimated by \citet{El-Badry2023a}.

\begin{table}
\caption{Predictions for the frequency of short-period black hole companions to solar-type stars, converted to $f_{\textrm{BH}; P<3}$ using the assumptions described in the text.}
\begin{tabular}{lr}
\hline
Paper &  $f_{\textrm{BH}; P<3}$\\
\hline
\citet{Masuda2019} & 65--70 $\times 10^{-6}$\\
\citet{Mashian2017} & 46 $\times 10^{-6}$\\
\citet{Wiktorowicz2020} & 0.93--19 $\times 10^{-6}$\\
\citet{Breivik2017} & 0.46--3.1 $\times 10^{-6}$\\
\citet{Wang2022} & 0.40 $\times 10^{-6}$\\
\citet{Chawla2022} & 0.22--0.37 $\times 10^{-6}$\\
\citet{Shikauchi2023} & 0.09--0.8 $\times 10^{-6}$\\
\citet{Shao2019} & 0.007--0.17 $\times 10^{-6}$\\
\hline
\end{tabular}
\label{tab:literature}
\end{table}

\subsection{Comparison to X-ray binaries}

In Fig.~\ref{fig:compare-literature}, we also plot the estimated frequency of low-mass X-ray binary systems (LMXBs) containing black holes, from \citet{Corral-Santana2016}.
These LMXBs consist of a black hole accreting matter from an FGKM-type donor star, and hence are accreting counterparts to the non-accreting systems for which we have searched.
As with BH-LCs, we have taken the estimated number of LMXBs in the Galaxy, and assumed that the spatial distribution of LMXBs in the Galaxy follows that of stars.
We have also divided the lower limit of the range from that work by a factor of two, as approximately half of LMXBs have main-sequence donors and the other half have evolved donors.

Mass transfer in LMXBs can be driven by one of two mechanisms: angular momentum loss driving the binary towards shorter periods, or the expansion of an evolving donor star.
In both scenarios, the binary spends some fraction of its lifetime as a non-accreting BH-LC before the onset of mass transfer.
Hence, it is reasonable to expect a population of non-accreting counterparts that may be similar or larger than the accreting population, with the ratio between the space densities of the two populations set by the durations of the corresponding evolutionary phases.
On the basis of the upper limit derived here, we can argue that the space density of non-accreting binary systems cannot be more than two orders of magnitude larger than that of the accreting systems.

\subsection{Potential false positives and false negatives}
\label{sec:positives}

Here, we discuss several potential sources of false positives (systems without black holes that are incorrectly selected as candidates) and false negatives (true black hole binaries that are missed) that may affect this survey or others that utilise the ellipsoidal selection methods discussed here.
As was noted previously, the dominant sources of false positives are binaries in which the primary star is somewhat inflated (in both selection methods) or close to Roche lobe-filling (in the {\tt qmin} selection method).
At short periods, contact binaries are an additional source of false positives.
Removal of these two sources of contamination from single-band photometry alone is not trivial, though \citet{Pesta2024} have recently put forward a method that may significantly reduce the number of contaminants selected.

Star spots constitute a source of both false positives and false negatives.
Stars in binary systems at these short periods are presumably tidally locked, and therefore rotating with spin periods of a few days.
Such fast-rotating stars are generally likely to have a high surface filling fraction of star spots \citep[e.g.][]{Cao2022}, and tidally locked star spots will not be removed by phase-folding the light curve on the orbital period.

Star spots at the anti-stellar points will have the effect of reducing the flux from the binary during one of the two light minima \citep[e.g.][]{Nagarajan2023}.
The {\sc beer} fit will interpret this as a larger $A_{\rm ell}$ with some additional reflection effect or gravitational darking, leading to an over-estimate of $q_{\rm min}$, which may cause the target to become a false positive.
Star spots in binary systems are preferentially found at the sub- and anti-stellar points on the stellar surfaces \citep[e.g.][]{Sethi2024}, which may increase the probability of false positives relative to negatives.
False positives introduced in this way are not a major concern, as they will be removed by spectroscopic follow-up.

The potential for false negatives due to spots on stars is more concerning.
Star spots on the leading or trailing face of one of the stars will reduce the flux during one of the maxima, which the {\sc beer} fit will interpret as a reduced $A_{\rm ell}$ combined with a Doppler beaming signature---this scenario is sometimes known as the O'Connell effect \citep{OConnell1951,Knote2022}.
In \citetalias{Green2023}, we applied a cut to remove any system in which the orbital amplitude was larger than the ellipsoidal ($a_1 > a_2$ in the notation used in that paper), on the grounds that this removed many non-binary pulsating stars from the sample and that the implied beaming signatures were not physical.
However, three recently-published, active K-dwarfs with high-mass white dwarf companions \citep{Tucker2024,Rowan2024b} were removed by this cut, because the O'Connell effect introduced by the star spots is larger than the ellipsoidal amplitude.
If not for this cut, all three would have been selected with $q_{\rm min} > 1$.

False negatives resembling those white dwarf binaries are naturally concerning for the results of this work, as they will reduce the selection efficiency $S$ discussed in Section~\ref{sec:upper-lim} and hence weaken the constraint on the space density of black hole binaries presented here.
A thorough incorporation of star spots into our injection-recovery simulations (Section~\ref{sec:simulations}) would require a number of assumptions about both the frequency and placement of star spots as a function of stellar temperature and spin period, which are not trivial to determine.
Although we have not attempted to quantify the effect of star spots on $S$, we do note that a reduction by a factor of $\gtrsim 2$ would be necessary to qualitatively change the results presented in the previous section.
We strongly recommend that future works adopt less strict cuts on the orbital amplitude in order to not remove binary systems with substantial O'Connell effects.

Also worth noting in this context, though not an issue for the survey presented here, are binary systems containing stripped giant stars (which have undergone historic Roche lobe overflow).
These are a source of false positives that may affect both photometric selection of candidates and spectroscopic follow-up \citep{Jayasinghe2021,El-Badry2022c,El-Badry2022d}.
In these systems, the stripped donor is substantially hotter and lower-mass than is possible for single-star evolutionary tracks, which can lead to an over-estimation of its mass and hence that of the (fainter, main-sequence) companion.
Such systems typically have orbital periods of $\approx 10$--20\,days, putting them outside the period range that the survey in this work was sensitive to, but would be relevant for any ellipsoidal search considering giant or sub-giant primary stars.



\section{Conclusions and outlook}
\label{sec:conclusions}

We have derived and presented an upper limit on the frequency of short-period black hole companions to solar-type stars, based on the \textit{TESS} light curves of 4.7 million stars. This involved the construction of an initial candidate sample, and the ruling out of the $\sim 250$ most promising candidates within the sample.
Black holes with orbital periods $< 3$\,days can exist around a fraction no greater (to $2\sigma$ confidence) than $9.5 \times 10 ^{-6}$ of all AFGK-type stars.
If we restrict our period range of interest to orbital periods close to 1\,day, the upper limit can be tightened to $\lesssim 1 \times 10^{-6}$.
We note that this upper limit may be affected by false negatives due to star spots, and we make recommendations for how to reduce this issue in future work.

This upper limit is sufficiently tight to strongly rule out the most optimistic predictions in the literature \citep[e.g.][]{Masuda2019,Mashian2017}, and challenge intermediate predictions---such as \citet{Breivik2017} or \citet{Wiktorowicz2020}---at the 1--2\,$\sigma$ level.
More pessimistic predictions, including some of the most recent works, will require a much larger sample size.

The ellipsoidal selection method, which requires only light curves, remains the most efficient method of searching for BH-LCs at short periods, despite drawbacks (discussed in Section~\ref{sec:candidates}) that lead to a high false-positive rate.
Contamination arises from contact binaries (in both selection methods used here), and from systems in which the stars are close to Roche-lobe filling or somewhat inflated (in the \qmin\ method).
We have highlighted the extremely low completeness of the \mmmr\ selection method proposed by \citet{Gomel2021c}, which was also used to select the \textit{Gaia} ellipsoidal sample \citep{Gomel2023}.
Because of this, we have found that tighter constraints on the population can be found using the \qmin\ method, as long as false positives can be effectively removed with follow-up observations.
We would suggest that future works focus on the \qmin\ method.

If we wished to explore the existence of short-period BH-LCs to a depth necessary to test recent population models at the 2--3\,$\sigma$ level, we would require a photometric survey with some 30--100\,$\times$ more targets than processed here. 
Deeper reductions of the \textit{TESS} data, such as the \textit{TESS-Gaia} light curves, that have a limiting magnitude of 16 \citep{Han2023}, would increase the number of targets observed.
An alternative may be to turn to large-footprint surveys with deeper limiting magnitudes such as the Zwicky Transient Facility, although the absence of continuous coverage may introduce complications.
Expanding the search to use the \qmin\ method at \porb$\,\lesssim 1\,$day would also enable a deeper limit to be achieved from the same set of input targets (Fig.~\ref{fig:upperlim-2d}), at the expense of increasing the false positive rate by allowing contact binary systems into the sample.

In future projects, given the larger sample size and fainter targets, stricter methods to remove contaminants (especially contact binary systems) will be essential to keep follow-up costs manageable.
A promising approach is the principal component analysis method put forward by \citet{Pesta2024}, which is effective at removing  both contact and detached MS-MS contaminants.
SED fitting may also be a cheap way to remove MS-MS binary systems \citep[as is demonstrated by][]{Kapusta2023}, and reliable estimates of primary stellar masses and radii will also reduce contamination.
If these methods prove robust at filtering out the majority of false positives, the ellipsoidal method will continue to be a valuable tool for constraining the population of BH-LCs at short orbital periods.

\section*{Data Availability}

Tables 3 and 4 are only available in electronic form at the CDS via anonymous ftp to \url{cdsarc.u-strasbg.fr} (130.79.128.5) or via \url{http://cdsweb.u-strasbg.fr/cgi-bin/qcat?J/A+A/}.

\begin{acknowledgements}

We thank the anonymous reviewer for their feedback which has improved the quality of the manuscript.
We also thank Dominick Rowan for comments on a pre-print of this work.
We are grateful to the European Southern Observatory and Isaac Newton Group of Telescopes staff for the additional support provided for operations during the COVID-19 pandemic.

MJG and HWR acknowledge support from the European Research Council through ERC Advanced Grant No. 101054731. MJG and DM acknowledge further funding under the European Union's FP7 Programme, ERC Advanced Grant No. 833031.

This work is based in part on observations collected at the European Organisation for Astronomical Research in the Southern Hemisphere under ESO programme 108.226R.
The NTT is operated by ESO at Cerro La Silla.
INT observations were obtained under proposal I/2022A/13.
The INT is operated on the island of La Palma by the Isaac Newton Group of Telescopes in the Spanish Observatorio del Roque de los Muchachos of the Instituto de Astrof\'{i}sica de Canarias.
This paper includes data collected by the \textit{TESS} mission. Funding for the \textit{TESS} mission is provided by the NASA's Science Mission Directorate.
This work has made use of data from the European Space Agency (ESA) mission {\it Gaia} (\url{https://www.cosmos.esa.int/gaia}), processed by the {\it Gaia} Data Processing and Analysis Consortium (DPAC, \url{https://www.cosmos.esa.int/web/gaia/dpac/consortium}). Funding for the DPAC has been provided by national institutions, in particular the institutions participating in the {\it Gaia} Multilateral Agreement.

This work has made use of the software \textsc{topcat} and the \textsc{python} packages \textsc{numpy, matplotlib, scipy, astropy, emcee} \citep{Foreman-Mackey2013}, \textsc{aspired} \citep{mlam2023}, \textsc{sparta} \citep{Shahaf2020}, and \textsc{phoebe} \citep{Prsa2005}.

\end{acknowledgements}

\bibliographystyle{aa} 
\bibliography{refs} 

\end{document}